\documentclass[aps,prd,floatfix,superscriptaddress,preprintnumbers,amsmath,amssymb,showpacs,showkeys]{revtex4} 
\usepackage{amssymb}
\usepackage{amsmath}
\usepackage{amsfonts}
\usepackage{graphicx,epsfig}
\usepackage{tabularx}
\usepackage{bm}
\newcommand{\seq}{\begin{subequations}}
\newcommand{\sen}{\end{subequations}}
\newcommand{\eq}{\begin{eqnarray}}
\newcommand{\en}{\end{eqnarray}}

\def\shiftdown#1{#1\llap{\lower.04ex\hbox{#1}}}

\newcommand{\ra}{\rangle}
\newcommand{\la}{\langle}

\def\shiftdown#1{#1\llap{\lower.04ex\hbox{#1}}}

\begin{document}

\title{Role of scalar mesons in the beam asymmetry 
of $p \bar p$ and $\Lambda \bar \Lambda$ photoproduction at JLab} 

\author{Thomas Gutsche} 
\affiliation{Institut f\"ur Theoretische Physik,
Universit\"at T\"ubingen,
Kepler Center for Astro and Particle Physics,
Auf der Morgenstelle 14, D-72076 T\"ubingen, Germany}
\author{Serguei Kuleshov}
\affiliation{Departamento de F\'\i sica y Centro Cient\'\i fico 
Tecnol\'ogico de Valpara\'\i so (CCTVal), Universidad T\'ecnica
Federico Santa Mar\'\i a, Casilla 110-V, Valpara\'\i so, Chile}
\author{Valery E. Lyubovitskij}
\affiliation{Institut f\"ur Theoretische Physik,
Universit\"at T\"ubingen,
Kepler Center for Astro and Particle Physics,
Auf der Morgenstelle 14, D-72076 T\"ubingen, Germany}
\affiliation{Departamento de F\'\i sica y Centro Cient\'\i fico 
Tecnol\'ogico de Valpara\'\i so (CCTVal), Universidad T\'ecnica
Federico Santa Mar\'\i a, Casilla 110-V, Valpara\'\i so, Chile}
\affiliation{Department of Physics, Tomsk State University, 634050 Tomsk, 
Russia}
\affiliation{Laboratory of Particle Physics, Tomsk Polytechnic University, 634050 Tomsk, Russia}
\author{Igor T. Obukhovsky}
\affiliation{Institute of Nuclear Physics, Moscow
State University, 119991 Moscow, Russia} 

\today

\begin{abstract}

We suggest a description of the beam asymmetry
in $p \bar p$ and $\Lambda \bar \Lambda$ photoproduction off the proton 
$\vec\gamma + p \to p \bar p + p$  
and $\vec\gamma + p \to \Lambda \bar\Lambda + p$,    
 takes into account the contribution of the scalar mesons 
$f_0(1370)$, $f_0(1500)$, and $f_0(1710)$. These scalars are considered 
as mixed states of a glueball and nonstrange and strange quarkonia 
in the framework based on the use of effective hadronic Lagrangians. 
Present results can be used to guide the possible search for
this reaction by the GlueX Collaboration at JLab. 
Also, we did an estimate of contribution of heavier scalar meson states 
$f_0(2020)$, $f_0(2100)$, and $f_0(2200)$. 

\end{abstract}

               

\maketitle 

\section{Introduction}

In this paper, we investigate the beam asymmetry in 
the $p\bar p$ and $\Lambda \bar \Lambda$ photoproduction 
due to the possible contribution of scalar mesons. 
This reactions are relevant to the physical program of the 
GlueX Collaboration (Hall D) at JLab. Note that the GlueX
Collaboration recently reported~\cite{AlGhoul:2017nbp} 
measurements of the photon beam 
asymmetry for the $\pi^0$ and $\eta$ photoproduction 
$\vec{\gamma} p \to p \pi^0$ and $\vec{\gamma} p \to p \eta$  
using a 9 GeV linear-polarized, tagged photon beam incident 
on a liquid hydrogen target.
The asymmetries, measured as a function of the proton momentum transfer, 
possess greater precision than previous $\pi^0$ measurements and 
are the first measurements involving the $\eta$ meson in this energy regime. 
The results are compared with theoretical 
predictions~\cite{Laget2011,Mathieu2015,Nys2017,Donnachie2016} 
based on $t$--channel, 
quasiparticle exchange and constrain the axial-vector component of 
the neutral meson production mechanism in these models. 

In present manuscript, we consider gluonic excitations 
in the intermediate mesons through photoproduction reactions. 
When focusing on events without 
really observed mesons, the detection of the glueball or a glueball 
component in a hadron is significantly simplified. The 
glueball will be present in these processes via its mixing 
with nonstrange  and strange quarkonia 
components~\cite{Giacosa2005,Chatzis2011}. In particular, 
the scalar fields $f_1 = f_0(1370)$, $f_2 = f_0(1500)$, 
and $f_3 = f_0(1710)$ are considered as mixed states of the glueball $G$ 
and non strange ${\cal N}$ and strange 
$S$ quarkonia~\cite{Giacosa2005,Chatzis2011}
$f_i = B_{i1} {\cal N} + B_{i2} G + B_{i3} S\,, \ i=1,2,3$,   
where the $B_{ij}$ are elements of the $3 \times 3$ mixing matrix rotating 
bare states $({\cal N}, G, S)$ into the physical scalar mesons $f_i$. 
Therefore, the glueball $G$ 
component will appear in the couplings of scalar mesons with photon 
and vector (axial) mesons and in the scalar meson propagators, which are 
the basic blocks for the calculation of the baryon-antibaryon 
photoproduction in our approach (see Fig.~1). Regarding the coupling 
of scalar mesons with $p\bar p$ and $\Lambda\bar\Lambda$ pairs,  
we proceed as follows (see details in Appendix): \\
1) We neglect by the coupling of glueball component 
to $p\bar p$ and $\Lambda\bar\Lambda$. \\
2) In case of $p\bar p$ photoproduction, we neglect by the coupling of strange 
quarkonia with $p\bar p$ and suppose that $f_ip\bar p$ couplings are 
dominated by the coupling of the nonstrange component of $f_i$ to nucleons. \\
3) In case of $\Lambda\bar\Lambda$ photoproduction, we take into account 
the couplings of both nonstrange and strange components of $f_i$ to 
$\Lambda$ hyperons. 

\begin{figure}[htb]
\begin{center}
\epsfig{figure=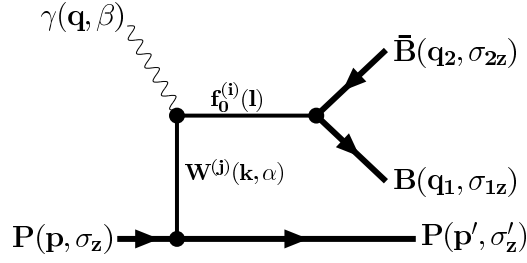,scale=0.75}
\caption{Relevant diagrams describing the contribution of intermediate
scalar mesons $f_0^{(i)}=$
$f_0(1370)$, $f_0(1500)$ and $f_0(1710)$ to the photoproduction
of the $B\bar B$ pair through the exchange of vector 
$V^{(j)}=\rho^0,\,\omega$ and axial-vector $A^{(j)}= b_1,\,h_1$ 
mesons (or the corresponding Reggeons). Here, $W = V, A$ and 
$B = p, \Lambda$.
} 
\label{f1}
\end{center}
\end{figure}

We start with definition of kinematics of the process of baryon-antibaryon 
photoproduction of the proton 
$\vec\gamma(q) + p(p)  \to p(p') + B(q_1) + \bar B(q_2)$ and introduce 
beam asymmetry:  
1) $p$, $p'$, $q$, $q_1$, and $q_2$ are the momenta of the initial 
and final protons, photon, produced baryon and antibaryon, respectively. 
2) Invariant Mandelstam variables $s$ (total energy), 
$t$ (the square momentum transferred to the target proton), and  
$s_2$ (the square of the invariant mass of the produced $B\bar B$ pair) 
are defined as 
\eq 
s &=& (p + q)^2 = (p' + q_1 + q_2)^2\,, \nonumber\\
t &=& k^2 = (p' - p)^2 = (q - q_1 -  q_2)^2\,, \nonumber\\
s_2 &=& (q_1 + q_2)^2\,. 
\en 
3) The asymmetry $A_{B\bar B}$, 
written according to the known Basel convention as
\eq 
A_{B\bar B}(t)=
\frac{d\sigma_{\bot}/d\Omega-d\sigma_{||}/d\Omega}
{d\sigma_{\bot}/d\Omega+d\sigma_{||}/d\Omega} = 
P_\gamma \Sigma\cos{2\varphi} 
\label{1}
\en
can be measured experimentally at JLab in a large interval of $t$. 
The numerator on the rhs of 
Eq.~(\ref{1}) is the difference of cross sections measured for linearly 
polarized photons, $\sigma_{||}$ for the polarization along the $x$ axis 
and $\sigma_{\bot}$ for the polarization along the $y$ axis, which are 
named the ``PARA'' and ``PERP'' orientations, respectively. 
The asymmetry $A_{B\bar B}(t)$ of Eq.~(\ref{1}) includes the factor 
$P_\gamma$ (the linear polarization of the initial photon beam),
 and thus the coefficient $\Sigma$ only can be considered as a beam asymmetry 
of the physical process.  
4) We use the laboratory (Lab) frame with the $z$ axis 
directed along the photon momentum $q^\mu=\{|\bm{q}|,0,0,|\bm{q}|\}$.   
The absolute value of the 3-vector of the transfer momentum 
$\bm{k}$ is expressed through $t$ and nucleon mass $m_N$ as 
$|\bm{k}|=\sqrt{-t\left(1-\frac{t}{4m_N^2}\right)}$. 
The beam asymmetry depends on the absolute value of $\bm{k}$ 
and the angles 
$\Omega = (\theta,\varphi)$ of $\bm{k}$ with 
respect to the photon 3-momentum $\bm{q}$ and the direction of 
the photon electric field $\bm{E}$ for the PARA variant of  
the polarization ($\bm{E}||\bm{x}$):
$k_x=|\bm{k}|\sin{\theta}\cos{\varphi}$, 
$k_y=|\bm{k}|\sin{\theta}\sin{\varphi}$, $k_z=|\bm{k}|\cos{\theta}$ with 
\eq
\cos{\theta} = 
\frac{1 + \frac{2m_N^2}{t} \frac{t - s_2}{s-m_N^2}} 
     {\sqrt{1 - \frac{4m_N^2}{t}}}  \,. 
\label{7}
\en 

In present paper, we consider theoretical predictions 
for the differential cross sections
\eq
\frac{d\sigma_{_{\rm P}}}{dt}=\int_{s_2^-}^{s_2^+}ds_2
\frac{d^2\sigma_{_{\rm P}}}{dt ds_2}\,, \quad 
\quad {\rm P} = \bot, || \, .
\label{2}
\en
As we mentioned before, the calculation is based on a model 
that takes into account the excitation of intermediate scalar mesons 
considered as mixed states of quarkonia and glueballs. 
The unpolarized cross section, which is given by the sum 
of both photon polarization cross sections with
\eq 
\frac{d^2\sigma}{dt ds_2} = 
\frac{d^2\sigma_{||}}{dtds_2}+\frac{d\sigma^2_{\bot}}{dt ds_2} =
\frac{1}{2} \, 
{\cal N}\left(\overline{|M_{||}|^2}+\overline{|M_{\bot}|^2}\right),
\quad \mbox{and} \quad  {\cal N}=
\frac{\alpha}{64\pi^2 (s-m_N^2)^2}\sqrt{\frac{s_2-4m_B^2}{s_2}},
\label{3}
\en
was considered in our recent work~\cite{Gutsche2016}. 
Here, $M_{||}$ and $M_{\bot}$ are the matrix 
elements for the PARA and PERP orientations of photoproduction, 
$\alpha = 1/137.036$ is the fine-structure constant, 
and $m_B$ is the mass of the produced baryon. 
The physical region of the reaction is constrained by the limits of 
the Chew-Low plot, defined by equations
\eq
s_2^- = 4m_B^2, \ 
s_2^+ = \frac{s-m_N^2}{2m_N^2} 
\left[\sqrt{t(t-4m_N^2)}+\frac{s+m_N^2}{s-m_N^2}\,t\right], \  
t^\pm = m_N^2-\frac{s-m_N^2}{2s}\biggl[\frac{s (s-s_2)}{s-m_N^2} 
- m_N^2 \mp \lambda^{1/2}(s,s_2,m_N^2)\biggr] 
\label{4}
\en
where $\lambda(x,y,z) = x^2 + y^2 + z^2 - 2xy - 2xz - 2yz$ 
is the K\"allen kinematical function.  
We have a characteristic value of 
\eq
t(s_{2max})=
m_N^2\left[2-\left(\frac{m_N}{\sqrt{s}}+\frac{\sqrt{s}}{m_N}\right)\right],
\quad 
s_{2max}=(\sqrt{s}-m_N)^2
\label{4a}
\en
that corresponds to the maximum condition 
$\frac{ds_2^+}{dt}\bigg\vert_{t=t(s_{2max})}=0$. 

\section{Formalism}

In this section we discuss the formalism for the calculation 
of the beam asymmetry in the process of the baryon-antibaryon 
photoproduction through the intermediate scalar meson based
on the models proposed and developed in 
Refs.~\cite{Giacosa2005,Chatzis2011,Gutsche2016}. 
The diagram in Fig.~\ref{f1} 
schematically represents the contribution of intermediate 
scalar mesons $f_1 = f_0(1370)$, $f_2 = f_0(1500)$ and $f_3 = f_0(1710)$ 
to the photoproduction 
of the $B_i\bar B_i$ (with $B_1 = p$, $B_2 = \Lambda$) 
pair through the exchange of vector 
$V_1 = \rho(770)$, $V_2 = \omega(782)$ with $J^{PC} = 1^{--}$ and axial-vector 
$A_1 = b_1(1235)$, $A_2 = h_1(1170)$ mesons with $J^{PC} = 1^{+-}$ 
(or the corresponding Reggeons).  

The full Lagrangian 
relevant for the  description of the $B\bar B$ photoproduction processes  
$\gamma + p \to B\bar B + p$ involving exchange by vector (axial) mesons 
in the $t$ channel and contribution of scalar mesons in the $s_2$ channel 
is given by a sum of free ${\cal L}_{\rm free}(x)$ and interaction 
${\cal L}_{\rm int}(x)$ 
Lagrangians~\cite{Giacosa2005,Chatzis2011,Gutsche2016}, 
\eq 
{\cal L}_{\rm full}(x) &=& {\cal L}_{\rm free}(x) +  {\cal L}_{\rm int}(x)\,,  
\nonumber\\  
{\cal L}_{\rm free}(x) &=& {\cal L}_F(x) + {\cal L}_f(x) 
+  {\cal L}_V(x) + {\cal L}_A(x) +  {\cal L}_B(x)\,, \\  
{\cal L}_{\rm int}(x)  &=&     
{\cal L}_{Vpp}(x) + {\cal L}_{App}(x) + 
{\cal L}_{fBB}(x) + 
{\cal L}_{fV\gamma}(x) + {\cal L}_{fA\gamma}(x) \,, \nonumber 
\en 
where ${\cal L}_F$, ${\cal L}_f$, ${\cal L}_V$, ${\cal L}_A$, and 
${\cal L}_B$ are free parts of electromagnetic field, scalar, 
vector, axial mesons, and baryons, respectively, 
\eq 
{\cal L}_F(x) &=& - \frac{1}{4} F_{\mu\nu}(x) F^{\mu\nu}(x)\,, \nonumber\\ 
{\cal L}_f(x) &=&   \frac{1}{2} \sum\limits_{i=1}^3 \Big[\partial_\mu f_i(x)
\partial^\mu f_i(x) - M_{f_i}^2 f_i^2(x) \Big]\,, \nonumber\\
{\cal L}_V(x) &=& - \frac{1}{2} \sum\limits_{i=1}^2 \Big[
\partial_\nu V_{i\mu}(x) \partial^\nu V_i^\mu(x) 
     - M_V^2 V_{i\mu}(x) V_i^\mu(x) \Big] \nonumber\\
{\cal L}_A(x) &=& - \frac{1}{2} \sum\limits_{i=1}^2 \Big[
\partial_\nu A_{i\mu}(x) \partial^\nu A_i^\mu(x) 
     - M_A^2 A_{i\mu}(x) A_i^\mu(x) \Big] \nonumber\\
{\cal L}_B(x) &=& \sum\limits_{i=1}^2\bar B(x) (i\not\!\partial - m_B) B(x)\,, 
\en 
and ${\cal L}_{Vpp}$, ${\cal L}_{App}$, ${\cal L}_{fV\gamma}$, 
${\cal L}_{fA\gamma}$, and ${\cal L}_{fBB}$  
are the interaction Lagrangians of vector and  axial mesons with protons, 
with scalar mesons and a photon, and scalar mesons with baryons,  
\eq 
{\cal L}_{Vpp}(x) &=& \sum\limits_{i=1}^2 \, \bar p(x) \, 
\biggl[ g_{V_ipp} \, V^\mu_i(x) \, \gamma_\mu 
+ \frac{f_{V_ipp}}{4 M_N} \, V^{\mu\nu}_i(x) \, 
\sigma_{\mu\nu} \biggr] p(x)\,, \nonumber\\ 
{\cal L}_{App}(x) &=& \sum\limits_{i=1}^2 \, \bar p(x) \, 
\biggl[ \frac{f_{A_ipp}}{4 M_N} \, A^{\mu\nu}_i(x) \, 
\sigma_{\mu\nu} \gamma_5 \biggr] p(x)\,, \nonumber\\ 
{\cal L}_{fBB}(x) 
&=& \sum\limits_{i=1}^3 \sum\limits_{k=1}^2 \, 
g_{f_iB_kB_k} \, f_i(x) \, \bar B_k(x) \, B_k(x) \,, \\
{\cal L}_{fV\gamma}(x) 
&=& \frac{e}{2} \, F_{\mu\nu}(x) \,  
\sum\limits_{i=1}^3 \sum\limits_{j=1}^2 \, 
g_{f_iV_j\gamma} \, f_i(x) \,  
V^{\mu\nu}_j(x)\,, \nonumber\\
{\cal L}_{fA\gamma}(x) 
&=& \frac{e}{4} \, e_{\mu\nu\alpha\beta }F^{\mu\nu}(x) \,  
\sum\limits_{i=1}^3 \sum\limits_{j=1}^2 \, 
g_{f_iA_j\gamma} \, f_i(x) \, A^{\alpha\beta}_j(x) \,.
\en 
Here we introduce the following notation: 
$F^{\mu\nu} = \partial^\mu {\cal A}^\nu - \partial^\nu {\cal A}^\mu$, 
$V^{\mu\nu} = \partial^\mu V^\nu - \partial^\nu V^\mu$,  
and $A^{\mu\nu} = \partial^\mu A^\nu  - \partial^\nu A^\mu$   
are the stress tensors of the electromagnetic field,   
vector, and axial mesons, respectively.

The scalar fields are considered as mixed states of the glueball $G$ 
and nonstrange $N$ and strange $S$ quarkonia~\cite{Giacosa2005,Chatzis2011}: 
$f_i = B_{i1} {\cal N} + B_{i2} G + B_{i3}$. 
The $B_{ij}$ are the elements of the mixing matrix rotating 
bare states $({\cal N}, G, S)$ into the physical scalar mesons 
$[f_0(1370), f_0(1500), f_0(1710)]$. 
In Refs.~\cite{Giacosa2005,Chatzis2011},  
we studied in detail different scenarios for the mixing of 
${\cal N}$, $G$, and $S$ states. Here, we proceed with the scenario 
fixed in Ref.~\cite{Chatzis2011} from a full analysis 
of strong $f_0$ decays and radiative decays of the $J/\psi$ with
the scalars in the final state: 
\eq\label{Bmatrix} 
B=\left(
\begin{array}{lll}
   0.75 & 0.60    & 0.26    \\
$-0.59$ & 0.80    & $-0.14$ \\
$-0.29$ & $-0.05$ & 0.95    \\ 
\end{array}
\right) \,.
\en
The coupling constants involving scalar mesons are given 
in terms of the matrix elements $B_{ij}$ and the effective couplings 
$c_f^s$ and $c_f^g$ of Ref.~\cite{Chatzis2011}:  
\eq 
g_{f_i\rho\gamma} = 3 g_{f_i\omega\gamma} 
\ = \ B_{i1} \, c_f^s + B_{i2} \, \sqrt{\frac{2}{3}} \, c_f^g \,.
\en   
The effective couplings $c_f^s = 1.592$ GeV$^{-1}$, 
and $c_f^g = 0.078$ GeV$^{-1}$ are 
fixed from data involving the scalar mesons $f_i$. 
In case of the $f_ipp$ couplings, we suppose that they are dominated by the 
coupling of the nonstrange component to the nucleon,  
\eq
g_{f_ipp} \simeq B_{i1} \, g_{{\cal N}pp} \,.
\en
The coupling $g_{{\cal N}pp}$ can be identified with the 
coupling of the nonstrange scalar $\sigma$ meson to nucleons,  
\eq\label{gsigmaNNcoupling}  
g_{{\cal N}pp} =  g_{\sigma pp} \simeq 5 \,.
\en  
In case of $f_i\Lambda\Lambda$ couplings, we take into account 
the coupling of both nonstrange and strange components to the $\Lambda$. 
We use the $SU(6)$ quark model relations in order to 
derive $f_i\Lambda\Lambda$ couplings. 

The invariant matrix element corresponding to the diagram 
in Fig.~\ref{f1} reads
\eq 
M^{V(ijk)\lambda}_{\rm inv}&=&
G^{V(ijk)}_{\rm eff} \, D^{(i)}_{f}(s_2) \, D^{(j)}_{V}(t) \, 
\Big[q^\alpha k^\mu - g^{\alpha\mu} k\cdot q\Big] 
\, \bar u_{B_k}(q_1,\sigma_{1z}) \, v_{B_k}(q_2,\sigma_{2z})\nonumber\\
&\times&\bar u_p(p^\prime,\sigma_z^\prime)
\left[\gamma_\alpha g_{V_jpp} + (k_\alpha - \gamma_\alpha\!\!\not k)
\frac{f_{V_jpp}}{2m_N}\right]u_p(p,\sigma_z) \, \epsilon^\lambda_\mu (q) 
\label{8}
\en
in the case of vector ($W=V$) meson exchange and
\eq 
M^{A(ijk)\lambda}_{\rm inv}&=&
G^{A(ijk)}_{\rm eff} \, D^{(i)}_{f}(s_2) \, D^{(j)}_{A}(t) \, 
\varepsilon^{\mu\nu\alpha\beta} q_\alpha k_\beta 
\, \bar u_{B_k}(q_1,\sigma_{1z}) \, v_{B_k}(q_2,\sigma_{2z})\nonumber\\
&\times&\bar u_p(p^\prime,\sigma_z^\prime)
\left[\gamma_\nu\gamma_5 \!\!\not k 
\frac{f_{A_jPP}}{2m_N} \right]u_p(p,\sigma_z) \, \epsilon^\lambda_\mu (q) 
\label{8a}
\en
in the case of axial-vector ($W=A$) meson exchange. 
The indices $i=1,2,3$; $j=1,2$, and $k=1,2$  
correspond to the summation over scalar [$f_1 = f_0(1370)$, 
$f_2 = f_0(1500)$, $f_3 = f_0(1710)$] and vector (axial-vector) 
[$V_1 = \rho^0$, $V_2 = \omega$, $A_1 = b_1$, $A_2 = h_1$]
mesons, and baryons [$B_1 = p$, $B_2 = \Lambda$], respectively. 
Here, $\bar u_{B_k}$ and $v_{B_k}$ are the spinors denoting the 
produced baryon and antibaryon; $\bar u_p$ and $u_p$ are 
the spinors denoting the final and initial proton; 
$\lambda=\pm$1 is the photon helicity; $\sigma_z$ 
is the baryon spin projection on the $z$ axis; 
$D^{(i)}_{f}(s_2)$ and  $D_{V(A)}^{(j)}(t)$ are 
the scalar and vector (axial-vector) meson propagators, respectively, 
including their resonance parts,  
\eq 
D^{(i)}_{f}(s_2) &=& \frac{1}{m_{f_i}^2 - s_2 - i m_{f_i} \, \Gamma_{f_i}} 
\,,\nonumber\\ 
D^{(j)}_{V(A)}(t) &=& \frac{1}{m_{V_j(A_j)}^2 - t - i m_{V_j(A_j)} 
\Gamma_{V_j(A_j)}} 
\,,
\en 
where a set of masses and the widths of scalar mesons,  
\eq 
   M_{f_1} = 1.432 \ {\rm GeV}\,, \quad
   M_{f_2} = 1.510 \ {\rm GeV}\,, \quad 
   M_{f_2} = 1.720 \ {\rm GeV}  
\en
and 
\eq 
   \Gamma_{f_1} = 350 \, {\rm MeV}\,, \quad 
   \Gamma_{f_2} = 109 \, {\rm MeV}\,, \quad 
   \Gamma_{f_3} = 135 \, {\rm MeV}
\en 
is the prediction of our model (see Refs.~\cite{Giacosa2005,Chatzis2011}), 
while for vector and axial mesons, 
we use cental values of data~\cite{PDG:2016}, 
\eq
& &\Gamma_{\rho} = 149.1\, {\rm MeV}\,, \quad 
   \Gamma_{\omega} = 8.49 \, {\rm MeV}\,, \nonumber\\  
& &\Gamma_{b_1} = 142\, {\rm MeV}\,, \quad 
   \Gamma_{h_1} = 360 \, {\rm MeV}\,. 
\en 
$G^{V(ijk)}_{\rm eff}(t,s_2)$ and $G^{A(ijk)}_{\rm eff}(t,s_2)$
are effective vertices, which are products of 
$fBB$, $fV\gamma$ and $fBB$, $fA\gamma$ phenomenological form factors, 
respectively, 
\eq 
G^{V(ijk)}_{\rm eff}(t,s_2) &=& g_{f_iB_kB_k}(s_2) \  
g_{f_iV_j\gamma}(t) \,, \nonumber\\
G^{A(ijk)}_{\rm eff}(t,s_2) &=& g_{f_iB_kB_k}(s_2) \  
g_{f_iA_j\gamma}(t) 
\en 
In Ref.~\cite{Gutsche2016} we dropped 
the $s_2$ and $t$ dependence of the corresponding form factors. 
However, 
in accordance with quark counting 
rules~\cite{Matveev:1973ra,Brodsky:1973kr,Lepage:1980fj},  
the form factors $g_{fBB}(s_2)$ and $g_{fV(A)\gamma}(t)$ 
should scale at large $s_2$ and $t$ as 
\eq 
& &g_{Vpp}(t) \sim \frac{1}{t^2}\,, \quad 
   f_{Vpp}(t) \sim \frac{1}{t^3}\,, \quad 
   f_{App}(t) \sim \frac{1}{t^3}\,, \nonumber\\
& &g_{fV(A)\gamma}(t) \sim \frac{1}{t}\,, \quad 
g_{fBB}(s_2) \sim \frac{1}{s_2^2}\,, 
\en 
These scalings following from the scaling results for the differential 
cross sections of 
the $p \bar p$ and $\Lambda \bar\Lambda$ pair production are consistent with 
the leading-twist quark fixed-angle counting
rules~\cite{Matveev:1973ra,Brodsky:1973kr,Lepage:1980fj}, 
\eq 
\frac{d\sigma}{dt}(A+ B \to C +D) \propto 
F(\theta_{\rm CM} )/s^{N-2}\,,
\en  
where $N= N_A + N_B + N_C + N_D $ is the total twist or
number of elementary constituents ($N_A = 1$ for the photon, 
$N_B = 3$ for the initial proton, $N_C = 6$ for the produced $B\bar B$ pair, 
and $N_D = 3$ for the final proton. In our case, we get $N - 2 = 11$. 
When we calculate the matrix element squared 
contributing to the differential cross section [see Eqs.~(\ref{MinvV2}] 
and~(\ref{MinvA2}) below), we find the product of $V(A)pp$, $fV(A)\gamma$,  
and $fBB$ form factors should scale as $1/t^3 \cdot 1/s_2^2$. 
Because of  $\rho(\omega)-\gamma$ universality,  
the Dirac and Pauli $V(A)pp$ form factors should scale 
as $1/t^2$ and $1/t^3$, respectively, to the 
scaling of the Dirac and Pauli $\gamma pp$ 
form factors. $fV(A)\gamma$ should scale as $1/t$ as other meson-meson-photon 
form factors. Finally, we conclude that the $fBB$ form factors 
should scale as $1/s_2^2$.  

We model the momentum dependence of hadronic form factors 
as 
\eq 
g_{fBB}(s_2) &=& g_{fBB}(M_f^2) 
\biggl[\frac{\Lambda_f^2 + M_f^2}{\Lambda_f^2 + s_2}\biggr]^2\,, 
\nonumber\\
g_{fV(A)\gamma}(t) &=& g_{fV(A)\gamma}(M_{V(A)}^2)
\frac{\Lambda^2_{V(A)}}{\Lambda^2_{V(A)} + M_{V(A)}^2 -  t}\,, \nonumber\\
g_{Vpp}(t) &=& g_{Vpp}(M_V^2) 
\biggl[\frac{\Lambda^2_V}{\Lambda^2_V + M_V^2 -  t}\biggr]^2\,,\nonumber\\ 
f_{V(A)pp}(t) &=& f_{V(A)pp}(M_{V(A)}^2) 
\biggl[\frac{\Lambda^2_{V(A)}}{\Lambda^2_{V(A)} + M_{V(A)}^2 -  t}\biggr]^3\,,
\label{0}
\en 
where $\Lambda_V$, $\Lambda_A$, and $\Lambda_f$ are the cutoff parameters. 
In numerical calculations, we will use for simplicity the universal parameter 
for $\Lambda_V$ and $\Lambda_A$, $\Lambda=\Lambda_V=\Lambda_A$, and 
fix its square $\Lambda^2$ at 0.7~GeV$^2$, i.e., at the value  
at which where results of the Born approximation are close to the Regge 
approximation results. Also, for a comparison, we will study 
a sensitivity of the results for the $B\bar B$ photoproduction 
to a variation of $\Lambda^2$ from 0.7 to 2~GeV$^2$. 
For $\Lambda_f$, we choose $\Lambda_f = 2M_B$. 
For convenience, we normalize the form factors on the mass shell 
of scalar and vector (axial) mesons: $s_2 = M_f^2$ and $t = M_V^2$ 
for $g_{fBB}(s_2)$ and $g_{fV(A)\gamma}(t)$, 
$g_{Vpp}(t)$, $f_{V(A)pp}(t)$, respectively.  
The couplings $g_{fNN}(M_f^2)$ and $g_{fV\gamma}(M_V^2)$ 
have been fixed in our previous paper~\cite{Gutsche2016}: 
\eq 
& &g_{f_1\rho\gamma} = 3g_{f_1\omega\gamma} = 1.24 \, {\rm GeV}^{-1}\,, 
\nonumber\\
& &g_{f_2\rho\gamma} = 3g_{f_2\omega\gamma} = -0.90 \, {\rm GeV}^{-1}\,, 
\nonumber\\
& &g_{f_3\rho\gamma} = 3g_{f_3\omega\gamma} = -0.47 \, {\rm GeV}^{-1}\,, 
\nonumber\\
& &g_{f_1NN} =  3.75 \,, \quad 
   g_{f_2NN} = -2.95 \,, \quad 
   g_{f_3NN} = -1.45 \,.
\en 
For the coupling constants $\rho pp$ and $\omega pp$
($g_{\rho pp}$, $g_{\omega pp}$, $f_{\rho pp}$, $f_{\omega pp}$)
we consider two variants, as in Ref.~\cite{Gutsche2016},
variant I and variant II, which are
\eq
& &g_{\rho pp} = 2.3 \,, \,\, g_{\omega pp} = 3 \, g_{\rho pp}\,,\,\, 
f_{\rho pp}    = 3.66 \, g_{\rho pp}\,,\,\, f_{\omega pp} = -0.07 \, 
g_{\omega pp} 
\quad
(\mbox{variant \,I})\,,\nonumber\\
& &g_{\rho pp} =  3.4,\,\, g_{\omega pp}  = 15,\,\, 
   f_{\rho pp} = 20.7,\,\, f_{\omega pp} = 0 \hspace*{3.15cm}
(\mbox{variant \,II})\,.
\label{6}
\en 
In case of axial meson couplings, we take $b_1pp$ and $h_1pp$ couplings 
from Ref.~\cite{Gamberg:2001qc},  
\eq 
g_{b_1pp} = 8.83\,, \quad g_{h_1pp} = 3.06
\en 
and identify the $f_iA\gamma$ couplings with corresponding $f_iV\gamma$ 
couplings 
\eq 
g_{f_i\rho\gamma} = 3g_{f_i\omega\gamma} = 
g_{f_ib_1\gamma}  = 3g_{f_ih_1\gamma}\,.
\en
The couplings of scalar mesons with hyperons are fixed using $SU(6)$ 
quark model relations (see details in the Appendix): 
\eq
g_{f_1\Lambda\Lambda}=4.699\,,\,\, 
g_{f_2\Lambda\Lambda}=-3.445\,,\,\, 
g_{f_3\Lambda\Lambda}=1.908\,.
\en 

In both cases (Feynman propagators and Regge trajectories), the spin 
structure of the corresponding vertices are equivalent to each other, 
and thus we only have to calculate the vector (axial-vector) meson vertex. 
It is further sufficient to substitute the Regge trajectories for the 
scalar parts of the vector (axial-vector) meson propagators as
\eq 
\frac{1}{t-m_V^2}\,\to\, D_V(t) = \left(\frac{s}{s_0}\right)^{\alpha_V(t)-1}
(-\alpha_V^\prime)\,\Gamma(1 - \alpha_V(t))\frac{-1+e^{i\pi\alpha_V(t)}}{2}
\label{5}
\en
into the final expression, 
where $\alpha_V(t)=\alpha_{0V}+\alpha^\prime_Vt$.  
 In the case of a single Regge trajectory, 
the factors (\ref{5}) do not influence the value of the ratio 
(\ref{1}) because they cancel each other in the numerator and 
the denominator. But in the case of several trajectories, the ratio~(\ref{1}) 
can dramatically depend on the position of points $t_0$, where the Regge 
trajectory $\alpha_j(t)$ has a zero with $\alpha_j(t_{0j}) = 0$. 
For example, the zero point $t_{0\rho}\approx\, -0.6$ GeV$^2$ of
the $\rho$ meson trajectory does not coincide with the zero point
$|t_{0b_1}|\approx 0.01$ of the unnatural parity trajectory
[the unnatural parity $b_1(1235), h_1(1170)$ exchanges are allowed
for $f_0$ photoproduction because of charge parity conservation],
and in the region of $t$ close to $t_{0\rho}$ and $t_{0b_1}$, 
the beam asymmetry $\Sigma(t)$ for $f_0$ (or for $\pi^0,\,\eta$)
photoproduction can be represented in the lowest order of $(t-t_{0\rho})$
and $(t-t_{0b_1})$ by
\eq
\Sigma_{\rm Regge}(t)\approx\frac{c^2_\rho(t-t_{0\rho})^2
+c^2_{b_1}(t-t_{0b_1})^2+2c_\rho c_{b_1}(t-t_{0\rho})(t-t_{0b_1})}
{d^2_\rho(t-t_{0\rho})^2
+d^2_{b_1}(t-t_{0b_1})^2+2d_\rho d_{b_1}(t-t_{0\rho})(t-t_{0b_1})}\,,
\label{5a}
\en
where $c_\rho,\, c_{b_1}$ and $d_\rho,\, d_{b_1}$ are coefficients
at the first nonzero terms of Taylor series for Reggeons~(\ref{5})
involved in Eq.~(\ref{1}) (for the numerator and denominator, respectively).
These coefficients are defined by parameters of different mesons,
$\rho$ and $b_1$, and thus
$\frac{c_\rho}{d_\rho}\neq \frac{c_{b_1}}{d_{b_1}}$.
It is easily seen that $\Sigma_{\rm Regge}(t)$ will jump from the value
$\frac{c_\rho}{d_\rho}$ to the one of $\frac{c_{b_1}}{d_{b_1}}$ inside
a relatively small interval $-t_{0b_1}\leq -t\leq -t_{0\rho}$ that
disturbs the smooth behavior of this function.
As a result, the Regge model results in a large dip for
the beam asymmetry $\Sigma(t)$ in the region of
$-t\approx 0-\,$0.6 GeV$^2$ for the $\pi^0,\eta$
photoproduction~\cite{Laget:1996}. It may occur in the $f_0$
photoproduction as well.

Hence, we cannot only use the Feynman amplitudes for the evaluation 
of the asymmetry~(\ref{1}). 
The functions~(\ref{5}) also play an important role in
the formation of the $t$ dependence of $\Sigma$.
As in our recent work~\cite{Gutsche2016},  
we use two variants for the parameters of the Regge trajectories: 
$\alpha_{0\rho}=0.53$,
$\alpha_\rho^\prime =0.85$\,GeV$^{-2}$,  $\alpha_{0\omega}= 0.4$, 
$\alpha_\omega^\prime =0.85$\,GeV$^{-2}$, and $s_0 = 1$~GeV$^2$  
in the case of variant I and 
$\alpha_{0\rho}=0.55$,
$\alpha_\rho^\prime =0.8$\,GeV$^{-2}$, $\alpha_{0\omega}= 0.44$,
$\alpha_\omega^\prime =0.9$\,GeV$^{-2}$, and $s_0 = 1$ GeV$^2$  
for variant~II. 
Now, we add the unnatural parity trajectory with 
$\alpha_{0b_1} = 0.0676$, $\alpha_{0h_1} = 0.0418$, and  
$\alpha_{b_1}^\prime =\alpha_{h_1}^\prime=\,$0.7\, GeV$^{-2}$ 
in both variants. 

For the photon polarized along the $x$ axis, we define the 
polarization vector as $\epsilon^{||}_\mu(q)=
(-\epsilon^{+1}_\mu+\epsilon^{-1}_\mu)/\sqrt{2}=\{0,1,0,0\}$ and for the 
photon polarized along the $y$ axis we define it as 
$\epsilon^{\bot}_\mu(q)=
i(\epsilon^{+1}_\mu+\epsilon^{-1}_\mu)/\sqrt{2}=\{0,0,1,0\}$. 
The photon spin density matrices for such states have the simplest 
representation in terms of Lorentz indices $\mu,\nu$ in the Lab frame: 
\eq 
\rho^{||}_{\mu\nu}=
{\epsilon^{||}_{\nu}}^\dagger\epsilon^{||}_{\mu} 
= {\rm diag}(0,1,0,0)\,,
\qquad
\rho^{\bot}_{\mu\nu}=
{\epsilon^{\bot}_{\nu}}^\dagger\epsilon^{\bot}_{\mu} 
= {\rm diag}(0,0,1,0)\,. 
\label{9}
\en 
Using these expressions one can write the PARA and PERP parts of the cross 
section (\ref{3}) as
\begin{eqnarray}
\overline{|M_{\rm P}|^2}&=&
\frac{1}{2}\sum_{ij}\sum_{i^\prime j^\prime}
\sum_{\sigma_z\sigma^\prime_z}\sum_{\sigma_{1z}\sigma_{2z}}
{M^{(i^\prime j^\prime)}}^*(\sigma^\prime_z,\sigma_z,\sigma_{1z},\sigma_{2z};
\nu)
M^{(ij)}(\sigma^\prime_z,\sigma_z,\sigma_{1z},\sigma_{2z};\mu)
\rho^{\rm P}_{\mu \nu}\,, \,\,{\rm P}=||,\bot\,,
\label{10}
\end{eqnarray}
where we represent the lhs of Eq.~(8) as $M^{(ij)\lambda}_{\rm inv}=
M^{(ij)}(\sigma^\prime_z,\sigma_z,\sigma_{1z},\sigma_{2z};\mu)
\epsilon_\mu^\lambda$. The full invariant matrix element is the sum
over all scalar and vector mesons 
$M^\lambda_{\rm inv}=\sum_{ij}M^{(ij)\lambda}_{\rm inv}$.
Then, using the rhs of Eq.~(\ref{8}), one can obtain, after elementary 
calculations, the final expressions for the squared matrix 
elements~(\ref{10}):  
\eq 
\overline{|M_{\rm P}|^2}=4(s_2-4m_B^2)\sum_{ijk}\sum_{i^\prime j^\prime k^\prime} \, 
(W^{\mu\nu})^{ijk, i'j'k'} \, \rho^{\rm P}_{\mu\nu} \, .
\label{11}
\en
$W^{\mu\nu}$ is the hadronic tensor, which in the case of 
vector meson exchange factorizes as 
\eq 
W^{\mu\nu} = 
G^{V(i^\prime j^\prime k^\prime)}_{\rm eff}G^{V(ijk)}_{\rm eff}
D^{(i^\prime)}_{f}(s_2)D^{(i)}_{f}(s_2)
D^{(j^\prime)}_V(t)D^{(j)}_V(t)T^{\mu\nu, \perp}_{j j^\prime}\,,
\en 
where
\eq 
T^{\mu\nu, \perp}_{j j^\prime}=\frac{1}{2}
\biggl[(g_{V_jpp}+f_{V_jpp})\,(g_{V_{j'}PP}+f_{V_{j'}pp}) 
\, g^{\mu\nu}_\perp \, k^2 (k\cdot q)^2
\,+\,4\Big(g_{V_jpp} \, g_{V_{j'}pp}
-\frac{k^2}{4m_N^2}f_{V_jpp}\, f_{V_{j'}pp}\Big)
f^{\mu\nu}_\perp\biggr]\,.
\label{12}
\en 
Here, $g^{\mu\nu}_\perp$ and $f^{\mu\nu}_\perp$ are the tensors, 
which are  explicitly orthogonal to the photon momentum 
\eq 
  g^{\mu\nu}_\perp &=& 
  g^{\mu\nu} 
- \frac{k^\mu q^\nu}{k\cdot q} 
- \frac{k^\nu q^\mu}{k\cdot q} 
+ k^\mu k^\nu \frac{q^2}{(k\cdot q)^2}\,, \nonumber\\  
f^{\mu\nu}_\perp &=& 
n^\mu_\perp n^\nu_\perp \,, \quad 
n^\mu_\perp \ = \ \frac{1}{m_N^3} \, 
\Big[p^\mu \,(p'q)-p^{\prime\,\mu} \,(p^\prime q)\Big]\,, 
\en 
i.e., obey the transversity conditions
\eq
q_\mu g^{\mu\nu}_\perp = q_\nu g^{\mu\nu}_\perp \,, \quad 
q_\mu f^{\mu\nu}_\perp = q_\nu f^{\mu\nu}_\perp \,, \quad 
q_\mu n^\mu_\perp = 0\,. 
\en 
The final result can be written in terms of the Lorentz invariants 
$s,s_2,t$ by using equations $p^2={p^\prime}^2=m_N^2$, $p\cdot q=(s-m_N^2)/2$, 
$p^\prime \cdot q=(s+t-s_2-m_N^2)/2$, $p\cdot p^\prime =m_N^2-t/2$, and 
$p\cdot k=-t/2$. After summation over $\mu$ and $\nu$, one gets 
\eq\label{MinvV2}  
\left\{
\begin{array}{l}
\!\! \overline{|M^V_{||}|^2}\!\!\! \\
 { }\\
\!\! \overline{|M^V_{\bot}|^2}\!\!\! \\ 
\end{array}
\right\}
&=&(s_2-4m_B^2)\sum_{ijk}\sum_{i^\prime j^\prime k^\prime}
G_{\rm eff}^V(i^\prime j^\prime k^\prime,ijk)
D_{f_0 V}(i^\prime j^\prime,ij;s_2,t)
\nonumber\\
&\!\!\times&\!\!\!\!\!
\Biggl[2 \, \biggl(g_{V_jpp}(t)   \!+\!f_{V_jpp}(t)   \biggr) 
         \, \biggl(g_{V_{j'}pp}(t)\!+\!f_{V_{j'}pp}(t)\biggr) \, 
\biggl(\!\frac{-t}{4}\!\biggr) (s_2\!-\!t)^2 \nonumber\\
&+& 
\biggl(g_{V_jpp}(t)\,g_{V_{j'} pp}(t)\!-\!\frac{t}{4m_N^2}
 f_{V_jpp}(t)\,f_{V_{j'} pp}(t)\biggr) \, (s\!-\!m_N^2)^2|\bm{k}|^2 
{\sin}^2\theta \!\left\{
\begin{array}{l}
 {\!\!\cos}^2\varphi\!\!\! \\
 { }\\
 {\!\!\sin}^2\varphi\!\!\! \\ 
\end{array}
\right\}\!\Biggr]\!,
\label{13}
\en 
where $G^V_{\rm eff}(i^\prime j^\prime k^\prime,ijk)=
G^{V(i^\prime j^\prime k^\prime)}_{\rm eff}G^{V(ijk)}_{\rm eff}$ and
\eq
D_{f V(A)}(i^\prime j^\prime,ij;s_2,t)&=&
D^{(i^\prime) \dagger}_{f}(s_2) D^{(i)}_{f}(s_2)
D^{(j^\prime) \dagger}_V(t)D^{(j)}_V(t) \nonumber\\
&=& \frac{1}{(m_{f_i}^2 - s_2)^2 + m_{f_i}^2 \Gamma_{f_i}^2} \, \cdot \, 
\frac{1}{(m_{V_j(A_j)}^2 - t)^2 + m_{V_j(A_j)}^2 \Gamma_{V_j(A_j)}^2} \,.
\en
In the case of the diagram with the axial-vector meson exchange,  
one obtains an analogous expression with 
\eq\label{MinvA2} 
\left\{
\begin{array}{l}
 \overline{{|M^A_{||}|}^2}\!\!\! \\
 { }\\
 \overline{{|M^A_{\bot}|}^2}\!\!\! \\ 
\end{array}
\right\}
&=&(s_2-4m_B^2)\sum_{ijk}\sum_{i^\prime j^\prime k^\prime}
G^A_{\rm eff}(i^\prime j^\prime k^\prime,ijk)D_{fA}(i^\prime j^\prime,ij;s_2,t)
\nonumber\\
&\times&
\Biggl[
f_{A_{jpp}}(t) 
f_{A_{j'}pp}(t) 
(s-m_N^2)^2|\bm{k}|^2 
{\sin}^2\theta 
\left\{
\begin{array}{l}
 {\sin}^2\varphi\!\!\! \\
 { }\\
 {\cos}^2\varphi\!\!\! \\ 
\end{array}
\right\}\Biggr]\,.
\label{13a}
\en 
Note that ${\sin}^2\varphi$ and ${\cos}^2\varphi$ in the rhs  
column are exchanged when comparing the expression of Eq.~(\ref{13a}) to
the one of Eq.~(\ref{13}). 
Such a permutation corresponds to the change of the vertex 
$\gamma Vf_0$ with Lorentz structure 
$\Big[q^\alpha k^\mu - g^{\alpha\mu} k\cdot q\Big]$ to the  
$\gamma A f_0$ vertex with 
$\varepsilon^{\rho\sigma\alpha\mu} q_\rho k_\sigma $ in passing from the 
vector amplitude~(\ref{8}) to the axial-vector one~(\ref{8a}). 
The vertex 
$\gamma Vf_0$ generates the scalar product 
$\hat n \cdot \hat k = \sin\theta \, \cos\varphi$
(i.e., the factor $\sin^2\theta \, \cos^2\varphi$ in the PARA cross section), 
while the vertex 
$\gamma A f_0$ generates the vector product 
$\hat n \times \hat k \sim \sin\theta \, \sin\varphi$
(i.e., the factor $\sin^2\theta \,\sin^2\varphi$ in the PARA cross section), 
where $\hat n$ is the vector of photon polarization. 

The upper line in the lhs columns of Eqs.~(\ref{13})-(\ref{13a}) 
corresponds to the cross section for photon polarized along the $x$ axis, 
i.e., $\hat n = \hat x$. Thus,  
the contribution of the axial-vector exchange to the asymmetry~(\ref{1}) 
given by (PERP-PARA)$_{A}$ 
$\sim \cos^2\varphi - \sin^2\varphi = \cos 2\varphi$ has a negative 
sign when compared to the contribution of the vector exchange, 
(PERP-PARA)$_{V}$ 
$\sim \sin^2\varphi - \cos^2\varphi = - \cos 2\varphi$. 
It is also important to note that no interference occurs 
between the vector and axial-vector amplitudes~(\ref{8}) and~(\ref{8a}) 
in the spin average~(\ref{10}), and the substitution 
$M^{(ij)}\to M^{V(ij)}+M^{A(ij)}$ to Eq.~(\ref{10}) gives 
$ \overline{|M^V_{\rm P}|^2}+\overline{|M^A_{\rm P}|^2}$, $\rm P=||,\bot$.

Now the asymmetry~(\ref{1}) can be rewritten through the event yields 
$Y_{||}(t)$ and $Y_{\bot}(t)$, which are 
proportional to ${\cal N} \int\left(\overline{|M^V_{||}|^2}
+\overline{|M^A_{||}|^2}\right)ds_2$
and  ${\cal N} \int\left(\overline{|M^V_{\bot}|^2}
+\overline{|M^A_{\bot}|^2}\right)ds_2$, 
respectively. Using Eqs.~(\ref{13}) and~(\ref{13a}), one can obtain 
\eq 
A_{B\bar B}(t)=\frac{Y_{\bot}(t)-Y_{||}(t)}{Y_{\bot}(t)+Y_{||}(t)}=
\frac{I^V_{\rm num}(t)+I^A_{\rm num}(t)}{I^V_{\rm den}(t)
+I^A_{\rm den}(t)} 
\cos{2\varphi} = \Sigma(t) \cos{2\varphi}\,,
\label{14}
\en
where   
\eq 
I^V_{\rm num}(t)&=&-\int_{s^-_2}^{s_2^+}ds_2
\sqrt{\frac{s_2-4m_B^2}{s_2}}(s_2-4m_B^2)
\sum_{ij}\sum_{i^\prime j^\prime}G^V_{\rm eff}(i^\prime j^\prime k^\prime,ijk)
D_{f_0 V}(i^\prime j^\prime,ij;s_2,t)\nonumber\\
&\times&\biggl(g_{V_{j^\prime}pp}(t) \, g_{V_jpp}(t)
-\frac{t}{4m_N^2}f_{V_{j^\prime}pp}(t) \, f_{V_jpp}(t)\biggr) \, 
(s-m_N^2)^2|\bm{k}|^2
{\sin}^2\theta,
\label{15}
\en
\eq 
I^A_{\rm num}(t)&=&\int_{s^-_2}^{s_2^+}ds_2
\sqrt{\frac{s_2-4m_B^2}{s_2}}(s_2-4m_B^2)
\sum_{ij}\sum_{i^\prime j^\prime}G^A_{\rm eff}(i^\prime j^\prime k^\prime,ijk)
D_{f_0 A}(i^\prime j^\prime,ij;s_2,t)\nonumber\\
&\times&f_{A_{j^\prime}pp}(t) \, f_{A_jpp}(t) \, 
\left(\frac{-t}{4m_N^2}\right)(s-m_N^2)^2|\bm{k}|^2
{\sin}^2\theta,
\label{15a}
\en
\eq 
I^V_{\rm den}(t)&=&\int_{s_2^-}^{s_2^+}ds_2
\sqrt{\frac{s_2-4m_B^2}{s_2}}(s_2-4m_B^2)
\sum_{ij}\sum_{i^\prime j^\prime}G^V_{\rm eff}(i^\prime j^\prime k^\prime,ijk)
D_{f_0 V}(i^\prime j^\prime,ij;s_2,t)\nonumber\\
&\times&\biggl[2 \biggl(g_{V_jpp}+f_{V_jpp}\biggr) \, 
                 \biggl(g_{V_{j^\prime}pp}+f_{V_{j^\prime}pp}\biggr)  
\biggl(\frac{-t}{4}\biggr)(s_2-t)^2 \nonumber\\
&+&\biggl(g_{V_{j^\prime}pp}(t) \, g_{V_jpp}(t) - 
\frac{t}{4m_N^2}f_{V_{j^\prime}pp}(t) \, f_{V_jpp}(t)\biggr) \, (s-m_N^2)^2 \, 
|\bm{k}|^2{\sin}^2\theta\biggr]
\label{16}
\en
\eq 
I^A_{\rm den}(t)&=&\int_{s_2^-}^{s_2^+}ds_2
\sqrt{\frac{s_2-4m_B^2}{s_2}}(s_2-4m_B^2)
\sum_{ij}\sum_{i^\prime j^\prime}G^A_{\rm eff}(i^\prime j^\prime,ij)
D_{f_0 A}(i^\prime j^\prime,ij;s_2,t)\nonumber\\
&\times&f_{A_{j^\prime}pp}(t) \, f_{A_jpp}(t) \, \biggl(\frac{-t}{4m_N^2}
\biggr) (s-m_N^2)^2
|\bm{k}|^2{\sin}^2\theta 
\label{16a}
\en
Note that it is trivial to generalize Eq.~(\ref{14}) to the case of a
partially polarized photon beam ($P_\gamma\ne 1$) using the substitution
\eq 
Y_{||}(t)={\cal N}(1-P_\gamma\Sigma(t)\cos{2\phi}),\quad 
Y_{\bot}(t)={\cal N}(1+P_\gamma\Sigma(t)\cos{2\phi})\,.
\label{17}
\en 
Finally, we define the integrated beam asymmetry $\la \Sigma \ra$ as 
\eq 
\la \Sigma \ra = 
\frac{\int dt \,\big[ I^V_{\rm num}(t)+I^A_{\rm num}(t)\big]}
{\int dt \, \big[I^V_{\rm den}(t)+I^A_{\rm den}(t)\big]} \,,
\label{18}
\en 
where $I^A_{\rm num}$ defined in Eq.~(\ref{15a}) is negative, 
which should diminish the beam asymmetry
$\Sigma(t)$ generated by $\rho$ and $\omega$ exchange diagrams.

\section{Results}

We study the linearly polarized beam asymmetry $\Sigma(t)$ 
for the $p\bar p$ and $\Lambda\bar\Lambda$ photoproduction off the proton. 
We calculate the $t$ dependence of the beam asymmetry 
$\Sigma(t)$ for the photon energies $E_\gamma = 5$ and $9$ GeV (relevant for 
the JLab experiment) following Eqs.~(\ref{14})-(\ref{16a}) and using the 
photoproduction model recently developed in Ref.~\cite{Gutsche2016}.
The obtained results for the $p\bar p$ photoproduction are shown 
in Figs.~\ref{f2} and \ref{f3} for $E_\gamma=\,$5 and 9 GeV, respectively. 
The results for the $\Lambda\bar\Lambda$ 
photoproduction are shown in Fig.~\ref{f4} for $E_\gamma=\,$9 GeV. 

Note that at $E_\gamma=\,$5 GeV the
maximum value of $s_2$ in the $\Lambda\bar\Lambda$ channel defined by
Eq.~(\ref{4a}) is only 0.15 GeV higher than the $\Lambda\bar\Lambda$ threshold
value $4m_\Lambda^2$, and thus the $\Lambda\bar\Lambda$ cross section is
very small as compared to the one of $p\bar p$ production because of the 
small phase space.
For this reason, the $\Lambda\bar\Lambda$ cross section is not shown 
for $E_\gamma=\,$5 GeV.

One can see that in the considered energy interval the absolute value of 
asymmetry $\Sigma$ is increasing with the increasing of photon energy 
and approaching to $-1$ in the limit of large $s$ due to 
$\la \Sigma \ra = - 1 + O(1/s)$. 
For example, the contribution of vector 
meson exchange (Figs.~\ref{f2} and~\ref{f3}) to the integrated asymmetry
$\langle \Sigma \rangle$ is $-0.119$ at $E_\gamma = 5$ GeV, 
and it takes the value of $-0.509$ at $E_\gamma = 9$ GeV. 

The contribution of exchanged vector and axial-vector
mesons to the $p\bar p$ photoproduction is described in terms of a Regge pole 
model for two sets of effective parameters, coupling constants $G_{\rm eff}$ 
and the values of $\alpha_0,\,\alpha^\prime$, which are
characteristic of the Regge pole trajectories. It turns out, as seen in 
Figs.~\ref{f2}-\ref{f4}, that for the standard 
set used in meson exchange models (variant I) the calculated cross section 
is several times smaller than for the set usually used in the Regge 
approach (variant II). 
As was shown in Ref.~\cite{Gutsche2016}, the Born approximation results in
an overestimate of the cross section if one uses the vertices without 
form factors. Now, we show
that the insertion of form factors~(\ref{0}) restores the agreement between 
Regge model predictions and 
the description in terms of a modified Born approximation.

The beam asymmetry $\Sigma(t)$ does not depend on the explicit values of 
the effective parameters. 
The results of the calculations made in both the Regge and 
Born approximations are very close each other, if one neglects 
the axial-vector meson exchanges. 
The beam asymmetry $\Sigma(t)$ calculated in 
the Born approximation does practically not differ from the results 
of the Regge model calculations, 
except the region $-t \approx 0 - 0.6$ GeV$^2$, where the vector 
(axial-vector) meson trajectory
passes through zero [$\alpha_j(t_{0j})=0,\,\,j=V,A$]. Then, the 
denominator in Eq.~(\ref{18}) is close to zero. In such a situation,  
the behavior of $\Sigma(t)$ is determined by 
the approximation~(\ref{5a}), which predicts a nontrivial jump 
of $\Sigma(t)$ if $t_{0b1}\neq t_{0\rho}$.

Note that not only for the cases of vector and axial-vector Reggeon 
exchanges such jumps occur.
In the case of two different vector resonances, $\rho$ and $\omega$, 
the $t$ behavior of the
asymmetry $\Sigma(t)$ should also be disturbed by the same mechanism, if 
$t_{0\omega}\neq t_{0\rho}$. However, this can rather be considered 
as an artifact
of the Regge-pole approximation. 
For example, the zero points of $\rho$ and $\omega$ trajectories,
$t_{0\rho}$ and $t_{0\omega}$, for the widely used sets of parameters 
(e.g. for variant I or II) are very
close to each other because both sets practically correspond to the 
same trajectory (the trajectory 
of natural parity resonances). In practice, one can slightly 
change the parameters of the $\omega$
trajectory to obtain an exact equality $t_{0\omega}=t_{0\rho}$ 
(without any essential change in the
observables), and then the irregular behavior of $\Sigma(t)$ near $t_0$ 
disappears. Here, we use
such modified parameters for the $\omega$ trajectory in variant I 
($\alpha^\prime_{\omega\,mod}=\,$0.8355, $\alpha_{0\omega\,mod}=\,$0.4805) 
and for the $\rho$ trajectory in variant II  
($\alpha^\prime_{\rho\,mod}=\,$0.9143, $\alpha_{0\rho\,mod}=\,$0.4501), 
and thus there are no irregularities in the 
$\Sigma(t)$ behavior, when only the contributions of natural parity 
resonances ($V=\rho, \omega$) are taken into account [the lower curves 
"$V$" in Figs.~\ref{f2}(a), \ref{f3}(a) and~\ref{f4}(a)]. 
However, it would be impossible to cancel the irregularity of $\Sigma(t)$ 
near $t_0$, when one takes into account the
contribution of two {\it really different} trajectories 
(e.g., the trajectories for natural and unnatural parity
resonances: see curves "$V+A$" in Figs.~\ref{f2}-\ref{f4}), 
because in this case the zero
points of such trajectories should be different by physical terms. 

It is apparent that, in addition to the contribution of 
the $f_0(1370)$, $f_0(1500)$, and $f_0(1710)$ states in the observables of 
the baryon-antibaryon production, one should consider contribution of 
other meson resonances of positive charge parity, 
which are sufficient in the considered energy interval.
For example, poorly established scalar mesons $f_0(2020)$, $f_0(2100)$, 
and $f_0(2200)$ could give a large contribution in considered physical 
properties since their masses are close to the $p\bar p$ threshold.
Unfortunately, their coupling constants $g_{\gamma V f_i}$ and 
$g_{f_i NN}$ are poorly known.
Therefore, for a rough estimate of a role of such "background" processes,  
we calculate the asymmetry $\Sigma$ and the differential cross section 
$\frac{d\sigma}{dt}$, taking into account the contribution
of the $f_0(2020)$, $f_0(2100)$, and $f_0(2200)$ states for which we use 
the corresponding coupling constants defined for $f_0(1370)$, $f_0(1500)$, 
and $f_0(1710)$ states, respectively. 
We take masses and widths of the $f_0(2020)$, $f_0(2100)$, and $f_0(2200)$ 
from data~\cite{PDG:2016}: 
\eq 
& &M_{f_0(2020)} = 1.992 \, {\rm GeV}\,, \quad 
   M_{f_0(2100)} = 2.101 \, {\rm GeV}\,, \quad 
   M_{f_0(2200)} = 2.189 \, {\rm GeV}\,, \nonumber\\ 
& &\Gamma_{f_0(2020)} = 442  \, {\rm MeV}\,, \quad 
   \Gamma_{f_0(2010)} = 224  \, {\rm MeV}\,, \quad 
   \Gamma_{f_0(2200)} = 238 \, {\rm MeV}\,. 
\en 
The results, obtained within the Regge model and with taking into account 
$f_0(2020)$, $f_0(2100)$, and $f_0(2200)$ states, are shown 
in Figs.~\ref{f2}(a), \ref{f2}(b), \ref{f3}(a), and \ref{f3}(b).  
It is seen that additional intermediate mesons can significantly contribute to 
the cross section but they cannot significantly change the asymmetry $\Sigma$.

While in a Regge approximation the $t$ dependence of the cross section is 
fixed by the known parameters of Regge-pole trajectories, 
in a Born approximation the $t$ dependence is defined
by form factors which are poorly known. Moreover, a small variation of 
the cutoff $\Lambda$ in vertex form factors~(\ref{0}) leads 
to a large variation of the cross section as it is shown in 
Fig.~\ref{f5} for $\Lambda^2=\,$0.7, 0.8, 1, 1.2, and 2 GeV$^2.$
One can see that only for $\Lambda^2\approx\, 0.7-1$ GeV are the Born results 
close to the stable results of the Regge model, but even at a small 
enhancement of $\Lambda^2$ up to
2 GeV$^2$, the Born cross section increases in an order of magnitude.

One can estimate the role of the axial-vector mesons ($b_1$, $h_1$) 
in the formation of a beam asymmetry in $p\bar p$ ($\Lambda\bar\Lambda$) 
photoproduction comparing the Regge
results obtained without the $b_1+h_1$ contribution [the curves "$V$" 
in Figs.~\ref{f2}(a), \ref{f3}(a), and~\ref{f4}(a)]   
with the results that take into account all exchanges $\rho+\omega+b_1+h_1$ 
(the curves "$V+A$").
It is seen that adding the $b_1$ and $h_1$ contributions does considerably 
lower the asymmetry $\Sigma(t)$ [in accordance with the analytical 
results~(\ref{15}), (\ref{15a})] and only slightly increases
the differential cross section [see Figs.~\ref{f2}(a) and \ref{f2}(b)].  
This common qualitative conclusion does not depend on concrete values
of the poorly understood axial-vector meson coupling constants 
(following the evaluations made in 
Ref.~\cite{Laget:1996} on the basis of $\pi^0,\,\eta$ photoproduction, 
we use here the same values 
of couplings as for the corresponding vector meson coupling constants). 
Quantitatively, the effect of lowering the absolute value of beam asymmetry 
$| \Sigma(t) |$ 
through the $b_1+h_1$ Reggeon exchange depends on concrete values for 
the axial-vector coupling constants, and thus the new data on  
$p\bar p$ and $\Lambda\bar\Lambda$ photoproduction
would be very useful for their evaluation.
 
\begin{acknowledgments}

The authors thank Reinhard Schumacher for useful discussions.  
This work was supported
by the German Bundesministerium f\"ur Bildung und Forschung (BMBF)
under Project 05P2015 - ALICE at High Rate (BMBF-FSP 202):
``Jet- and fragmentation processes at ALICE and the parton structure        
of nuclei and structure of heavy hadrons'',
by the Basal Conicyt No. FB082, by CONICYT (Chile) PIA/Basal FB0821, 
by Fondecyt (Chile) Grant No. 1140471 and CONICYT (Chile) Grant No. ACT1406,
by Tomsk State University Competitiveness
Improvement Program and the Russian Federation program ``Nauka''
(Contract No. 0.1764.GZB.2017); 
by Tomsk Polytechnic University Competitiveness Enhancement 
Program (grant No. VIU-FTI-72/2017); by the Deutsche Forschungsgemeinschaft
(DFG Projects No. FA 67/42-1 and No. GU 267/3-1); and by the Russian Foundation
for Basic Research (Grant No. RFBR-DFG-a 16-52-12019). 

\end{acknowledgments}


\appendix


\section{Coupling constants for $\Lambda\bar\Lambda$ channel}

The scalar fields $f_i$ are considered as mixed states of the glueball $G$ 
and nonstrange ${\cal N}$ and 
strange $S$ quarkonia~\cite{Giacosa2005,Chatzis2011}
$f_i = B_{i1} {\cal N} + B_{i2} G + B_{i3} S$, 
where the $B_{ij}$ are elements of the mixing matrix rotating 
bare states $({\cal N}, G, S)$ into the physical scalar mesons 
$[f_0(1370), f_0(1500), f_0(1710)]$. 
In Refs.~\cite{Giacosa2005,Chatzis2011},  
we studied in detail different scenarios for the mixing of 
${\cal N}$, $G$, and $S$ states. Here, we proceed with the scenario 
fixed in Ref.~\cite{Chatzis2011} from a full analysis 
of strong $f_0$ decays and radiative decays of the $J/\psi$ with
the scalars in the final state: 
\eq\label{Bmatrix_app} 
B=\left(
\begin{array}{lll}
   0.75 & 0.60    & 0.26    \\
$-0.59$ & 0.80    & $-0.14$ \\
$-0.29$ & $-0.05$ & 0.95    \\ 
\end{array}
\right) \,.
\label{a1}
\en
The coupling constants involving scalar mesons are given 
in terms of the matrix elements $B_{ij}$ and the effective couplings 
 $c_f^s=\,$1.592 GeV$^{-1}$ and  $c_f^g=\,$0.078 GeV$^{-1}$
of Ref.~\cite{Chatzis2011} fixed from data involving  the scalar mesons $f_i$:
\eq 
g_{f_i\rho\gamma} &=& 3 g_{f_i\omega\gamma} 
\ = \ B_{i1} \, c_f^s + B_{i2} \, \sqrt{\frac{2}{3}} \, c_f^g \,.
\label{a2}
\en   
In case of the $f_iNN$ couplings, we suppose that they are dominated by the 
coupling of the nonstrange component to the nucleon 
$g_{f_iNN} \simeq B_{i1} \, g_{{\cal N}NN}$. 
The coupling $g_{{\cal N}NN}$ can be identified with the 
coupling of the nonstrange scalar $\sigma$ meson to nucleons,  
\eq\label{a3}
g_{{\cal N}NN} =  g_{\sigma NN} \simeq 5\,, 
\en  
which plays an important role in phenomenological approaches 
to the nucleon-nucleon potential generated by
meson exchange~\cite{Machleidt2000}. 

In the case of the couplings of scalar mesons with hyperons,  
we use $SU(6)$ quark model relations. 
The master formulas are:
\eq 
c_{{\cal N} p p} &=&    
\la p\uparrow | \, \sum\limits_{i=1}^3 I^{i}_N \, 
| p\uparrow\ra \,  \nonumber\\
c_{{\cal N}\Lambda\Lambda} &=&    
\la \Lambda\uparrow | \, \sum\limits_{i=1}^3 I^{i}_N \, 
| \Lambda\uparrow\ra \,  \nonumber\\
c_{S\Lambda\Lambda} &=&    
\la \Lambda\uparrow | \, \sum\limits_{i=1}^3 I^{i}_S \, 
| \Lambda\uparrow\ra \,,
\en   
where $I_N = {\rm diag}(\frac{1}{\sqrt{2}},\frac{1}{\sqrt{2}},0)$, 
$I_S = {\rm diag}(0,0,1)$.
Using the proton and $\Lambda$ hyperon SU(6) wave functions 
\eq 
| p\uparrow \ra &=& \frac{1}{\sqrt{2}} \biggl[ 
\frac{1}{6} (udu + duu - 2uud) 
(\uparrow \downarrow \uparrow 
+\downarrow \uparrow \uparrow 
- 2 \uparrow \uparrow \downarrow) \nonumber\\
&+& 
\frac{1}{2} (udu - duu) 
(\uparrow \downarrow \uparrow 
-\downarrow \uparrow \uparrow)  
\biggr) \,, \nonumber\\
| \Lambda\uparrow \ra &=& \frac{1}{\sqrt{12}} \biggl[ 
uds 
(\uparrow \downarrow \uparrow 
-\downarrow \uparrow \uparrow) 
+
dus  
(\downarrow \uparrow \uparrow 
-\uparrow\downarrow \uparrow) 
+
usd 
(\uparrow \uparrow \downarrow 
-\downarrow \uparrow \uparrow) \nonumber\\
&+&
sud 
(\uparrow \uparrow \downarrow 
-\uparrow \downarrow \uparrow) 
+
dsu 
(\downarrow \uparrow \uparrow 
-\uparrow \uparrow \downarrow) 
+
sdu 
(\uparrow \downarrow \uparrow 
-\uparrow \uparrow \downarrow) 
\biggr] 
\en 
gives 
\eq
c_{{\cal N} p p} = \frac{3}{\sqrt{2}}\,,\,\,c_{{\cal N}\Lambda\Lambda} = \sqrt{2}\,,\,\,
c_{S\Lambda\Lambda}= 1 \,. 
\en
Using our result for 
$g_{f_iNN} = B_{i1} \, g_{{\cal N} NN}$~\cite{Gutsche2016},   
we get 
$g_{f_i\Lambda\Lambda} = 
B_{i1} \, 
g_{{\cal N} \Lambda\Lambda} 
+ 
B_{i3} \, g_{S \Lambda\Lambda}$, 
where $g_{{\cal N} \Lambda\Lambda}$  
and $g_{S \Lambda\Lambda}$ are deduced from the ratios 
\eq 
\frac{g_{{\cal N} \Lambda\Lambda}}
{g_{{\cal N} pp}} = \frac{c_{{\cal N} \Lambda\Lambda}}{c_{{\cal N} pp}} = \frac{2}{3}\,, \quad
\frac{g_{S \Lambda\Lambda}}{g_{{\cal N} pp}} = 
\frac{c_{S \Lambda\Lambda}}{c_{{\cal N} pp}} = \frac{\sqrt{2}}{3} 
\en 
as
\eq 
g_{{\cal N} \Lambda\Lambda} = \frac{2}{3} g_{{\cal N} pp}\,, \quad
g_{S \Lambda\Lambda} = \frac{\sqrt{2}}{3} g_{{\cal N} pp}\,. 
\en
Using Eqs.~(\ref{a1})-(\ref{a3}) and the values of 
$g_{f_1 NN}=3.75$\,,\,\, 
$g_{f_2 NN}=-2.95$\,,\,\, 
and $g_{f_3 NN}=-1.45$\,, we get
\eq
g_{f_1\Lambda\Lambda}=4.699\,,\,\, 
g_{f_2\Lambda\Lambda}=-3.445\,,\,\, 
g_{f_3\Lambda\Lambda}=1.908\,.
\en 

\newpage

\begin{figure}
\begin{center}
\epsfig{figure=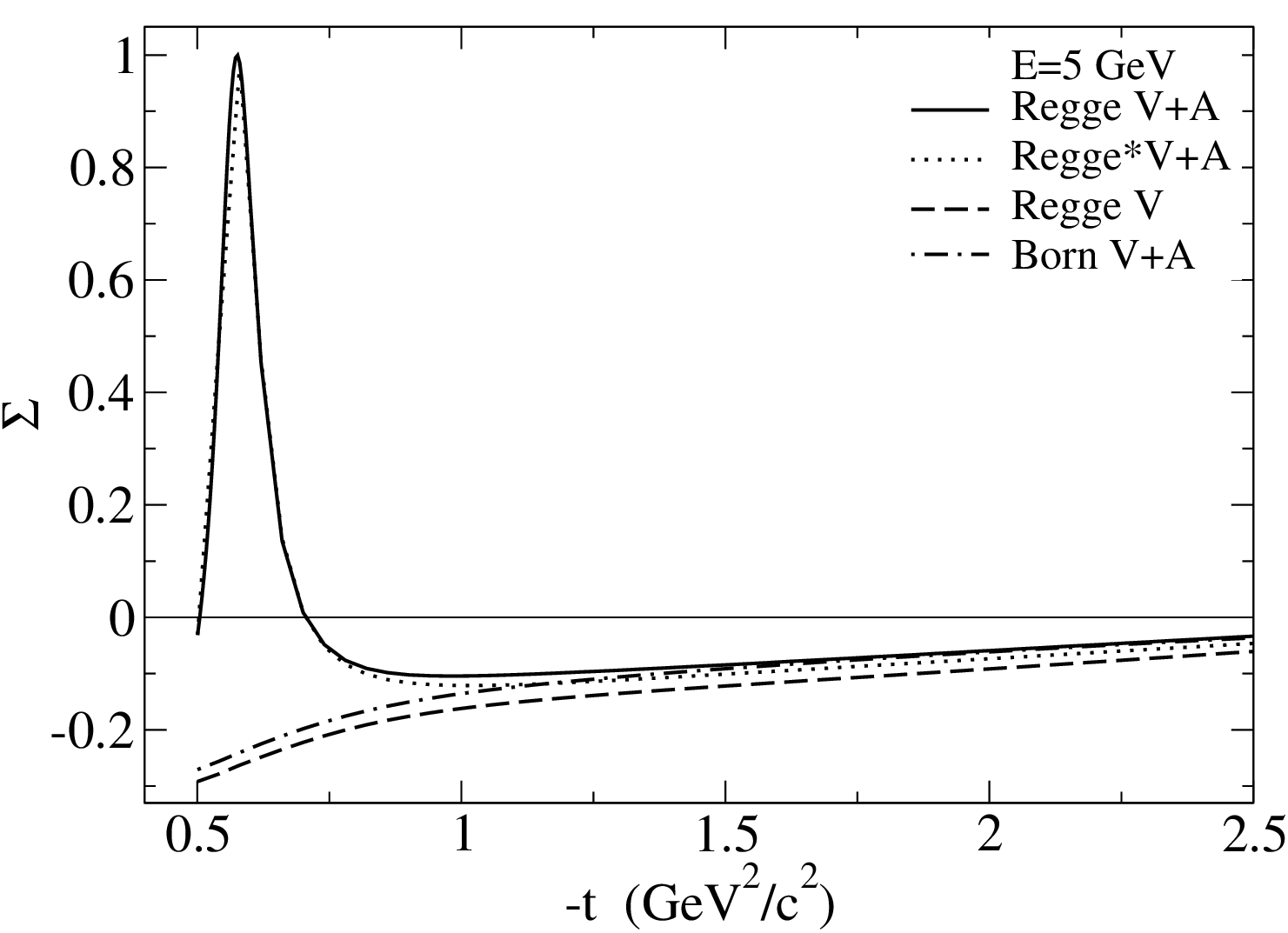,scale=0.2}
\quad
\epsfig{figure=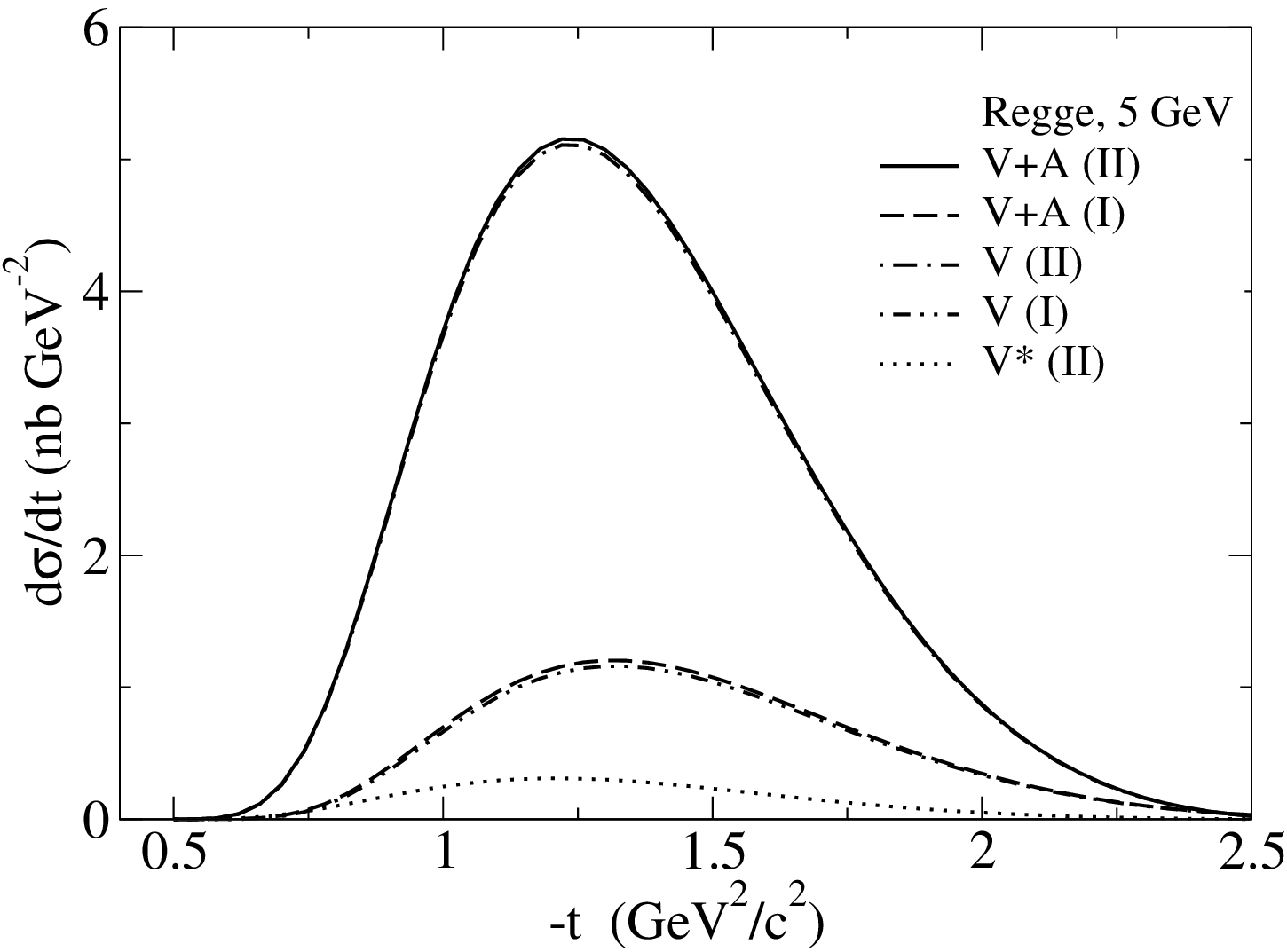,scale=0.2}
\quad
\epsfig{figure=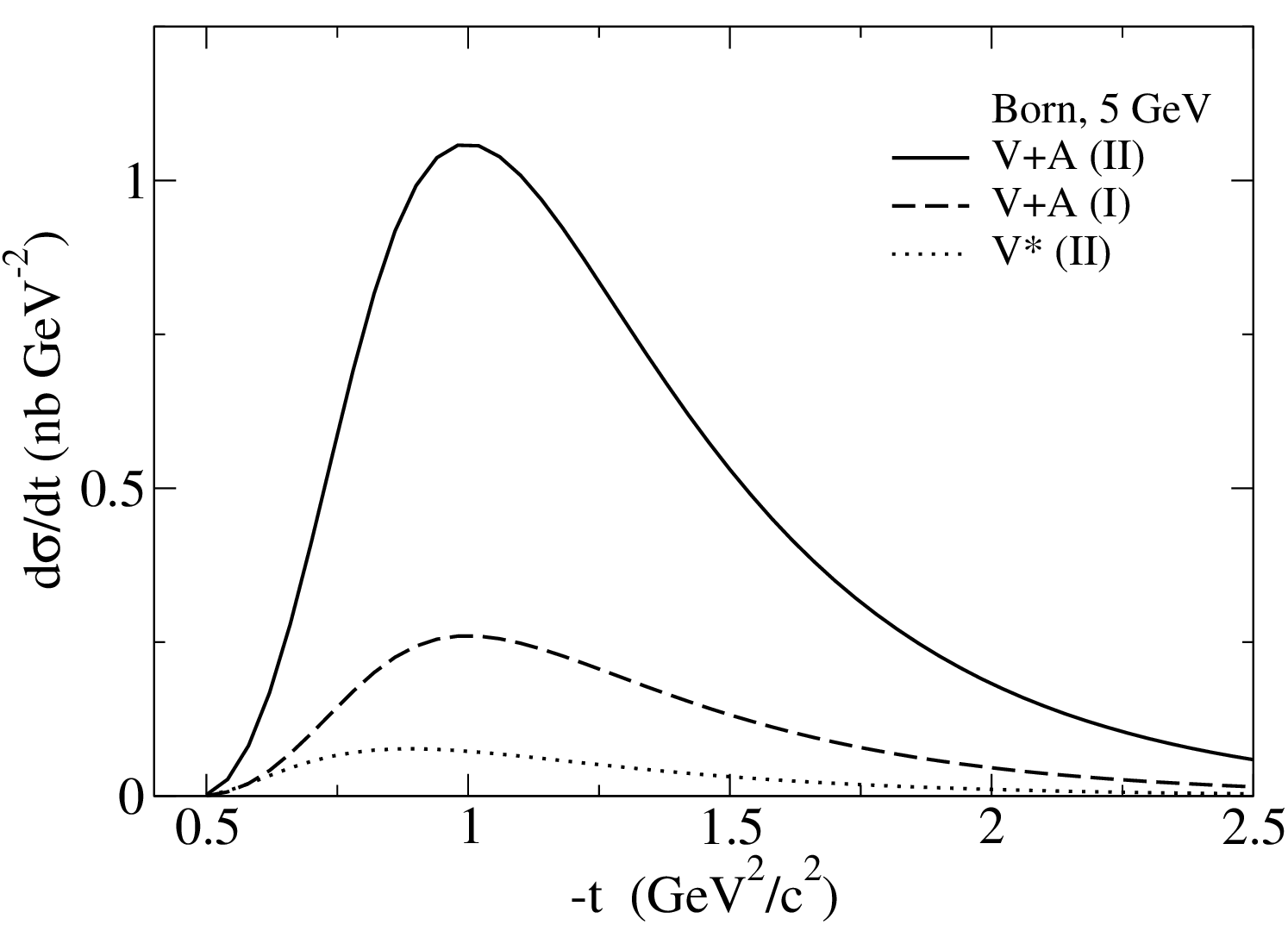,scale=0.2}

$(a)$\hspace{55mm}$(b)$\hspace{55mm}$(c)$
\vspace*{-.25cm}
\caption{The $p\bar p$ photoproduction off the proton, $E_\gamma=\,$5 GeV:  
($a$) beam asymmetry $\Sigma_{p\bar p}$, 
($b$) $d\sigma_{p\bar p}/dt$ in the Regge-pole approximation, 
($c$) $d\sigma_{p\bar p}/dt$ in the Born approximation.
In panel $a$,  
the lower two curves (without a peak at $-t\approx\,$0.6 GeV$^2$) 
correspond to the vector meson exchange ($V=\rho+\omega$) in the Born 
(dotted-dashed curve) and Regge-pole (dashed curve) approximations. 
The upper two curves are obtained for the sum of vector and axial-vector 
meson exchanges ($V+A=\rho+\omega+b_1+h_1$) for the Regge-pole approximation
with taking into account six (solid curve) and three (pointed curve) 
intermediate scalar mesons, respectively. 
In panel $b$, results for two sets of effective parameters are presented: 
the lower two curves -- 
variant I (dashed for $V+A$ and two-pointed dashed for $V$ exchanges) 
and the upper two curves, variant  II (solid for $V+A$ and dotted-dashed 
for $V$ exchanges). 
In panel $c$, the same notations for the curves are used as in panel $b$. 
Dotted curves in panels $a$, $b$, and $c$ show the results 
obtained with taking into account the contribution of only 
three intermediate scalar mesons, 
$f_0(1370)$, $f_0(1500)$, and $f_0(1710)$. 
The rest takes into account also the contribution 
of $f_0(2020)$, $f_0(2100)$, and $f_0(2200)$.}   
\label{f2}
\end{center}

\begin{center}
\vspace*{.15cm}
\epsfig{figure=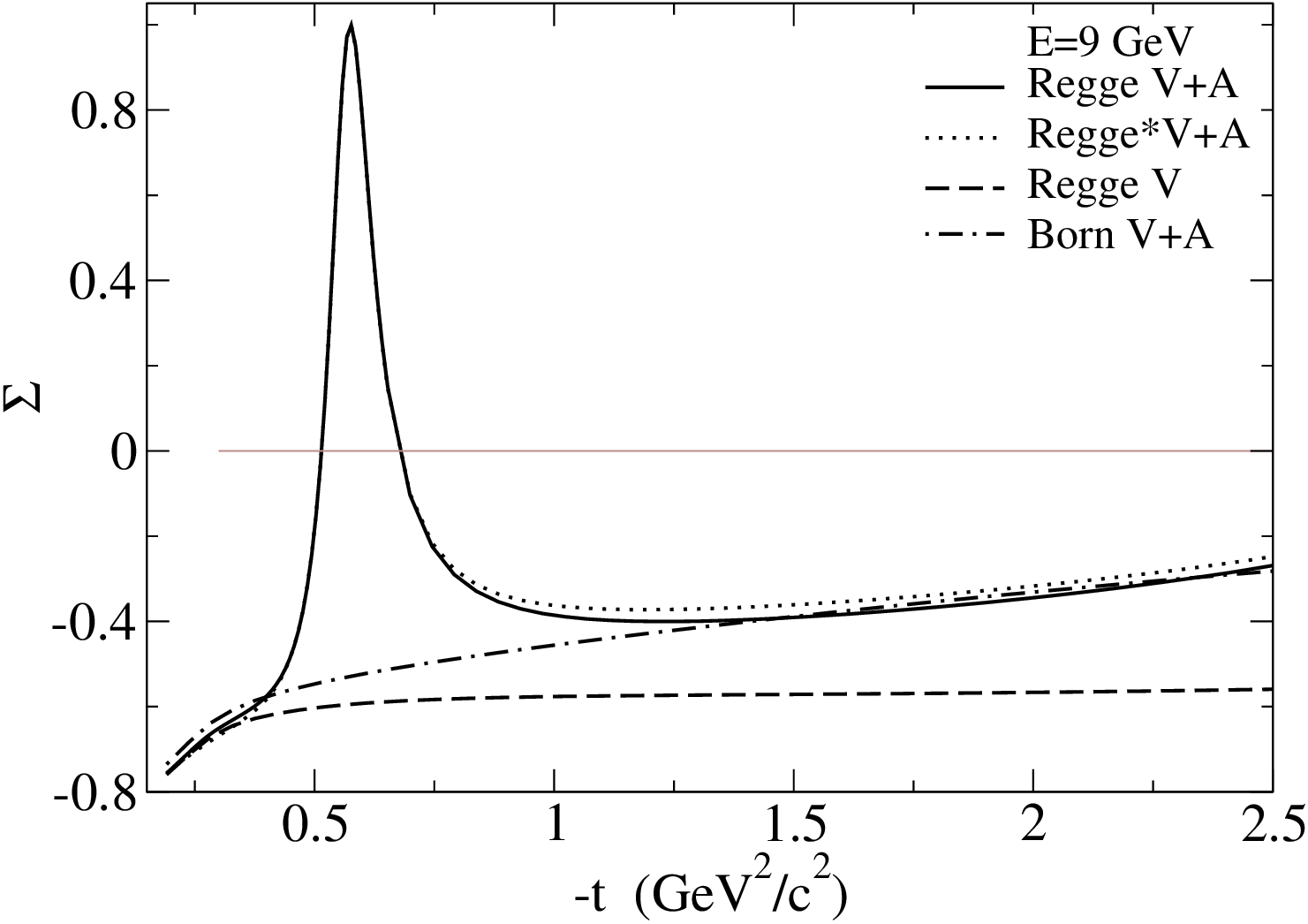,scale=0.2}
\quad 
\epsfig{figure=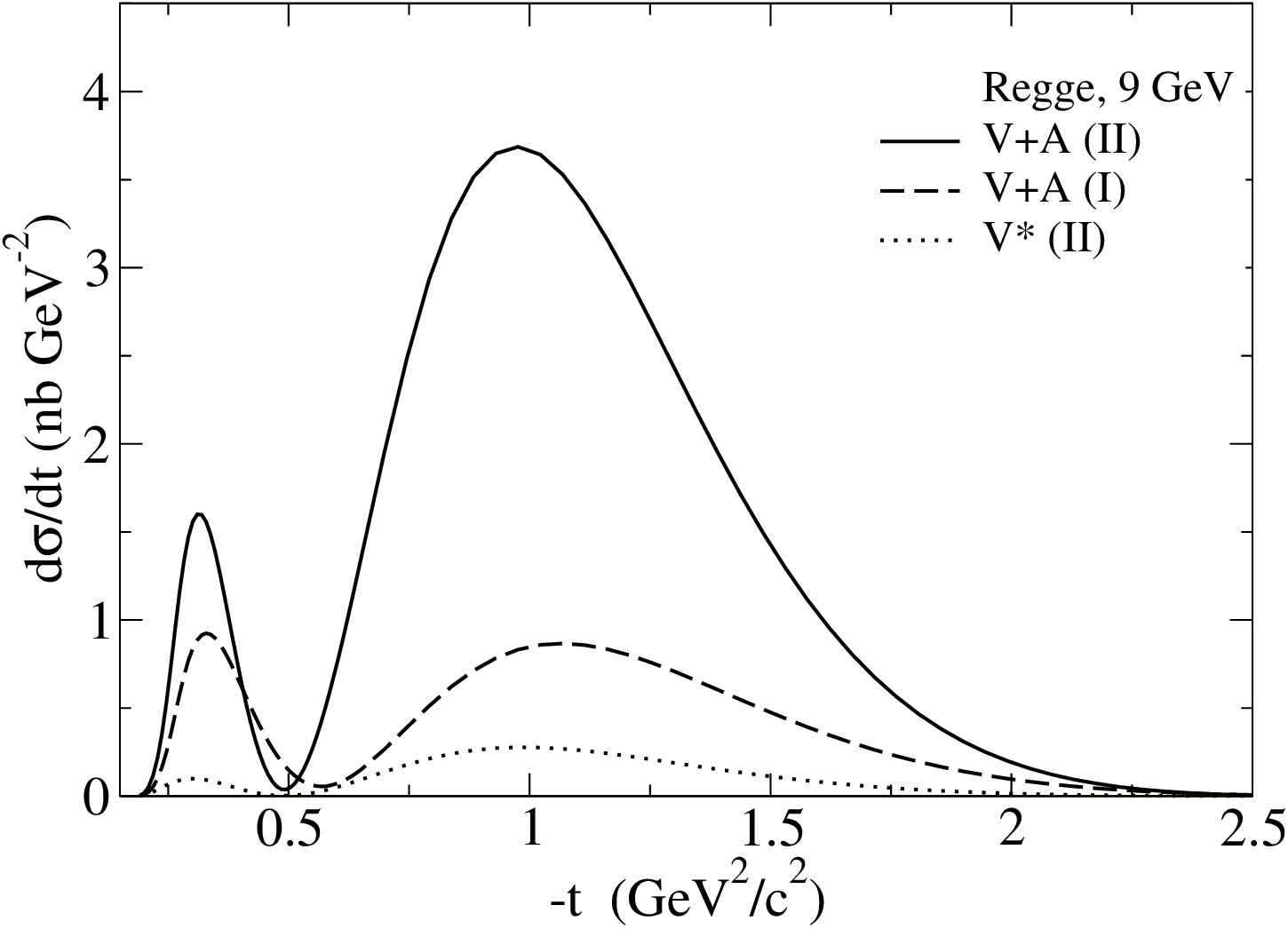,scale=0.2}
\quad 
\epsfig{figure=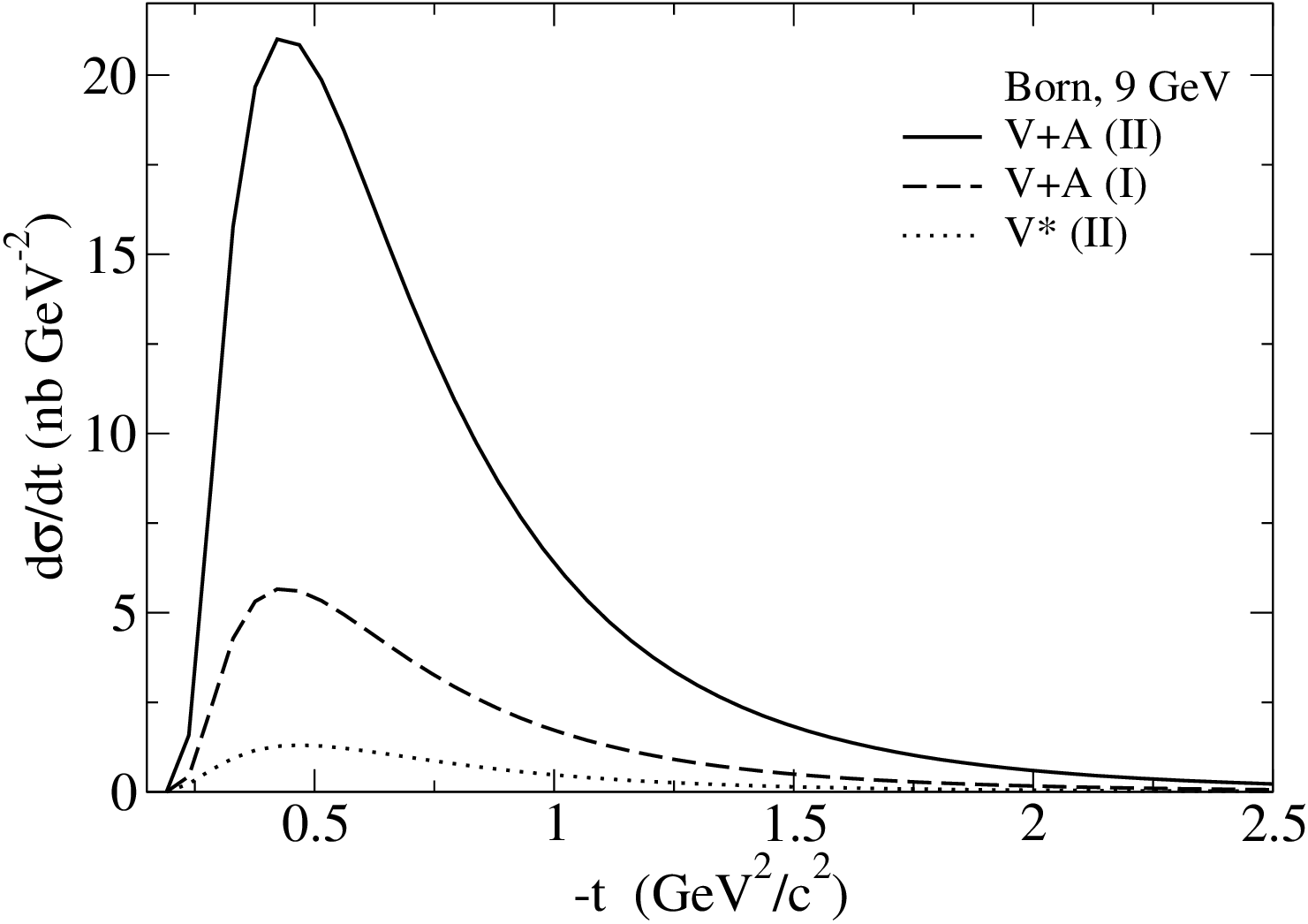,scale=0.2}

$(a)$\hspace{55mm}$(b)$\hspace{55mm}$(c)$
\vspace*{-.25cm}
\caption{The $p\bar p$ photoproduction for $E_\gamma=\,$9 GeV. 
The same content of panels as in Fig.~\ref{f2}.}

\label{f3}
\end{center}

\begin{center}
\vspace*{.15cm}
\epsfig{figure=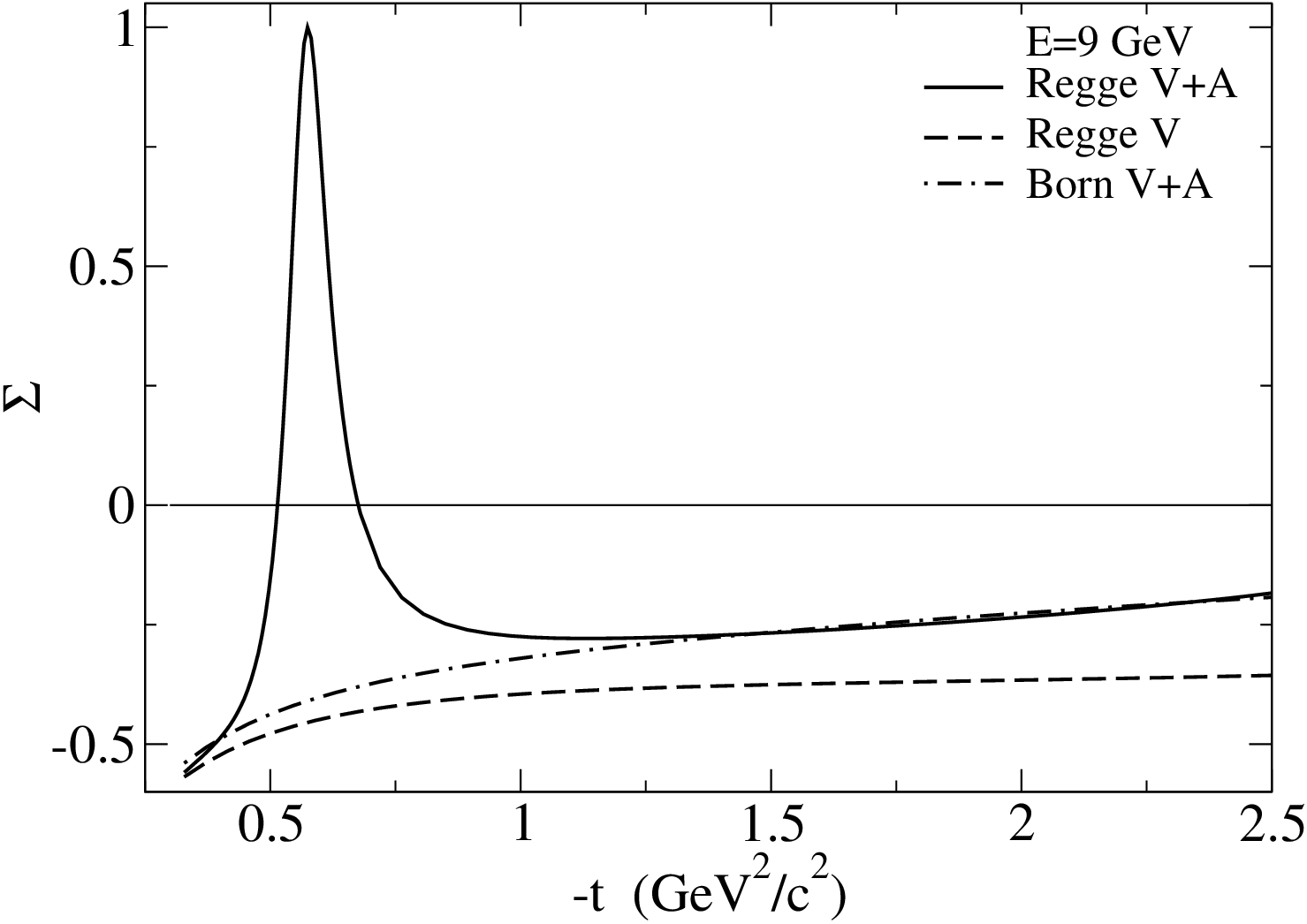,scale=0.2}
\quad 
\epsfig{figure=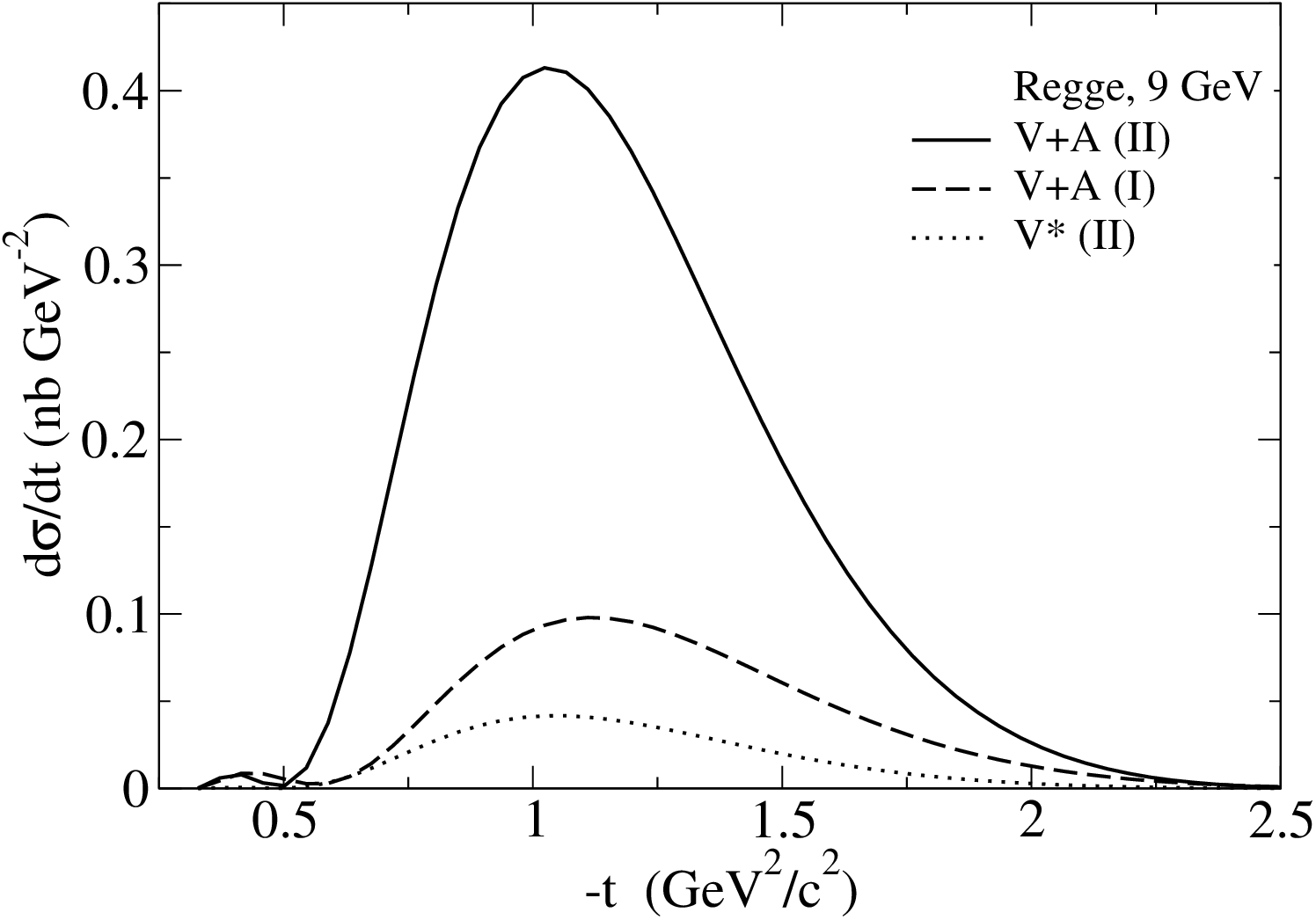,scale=0.2}
\quad 
\epsfig{figure=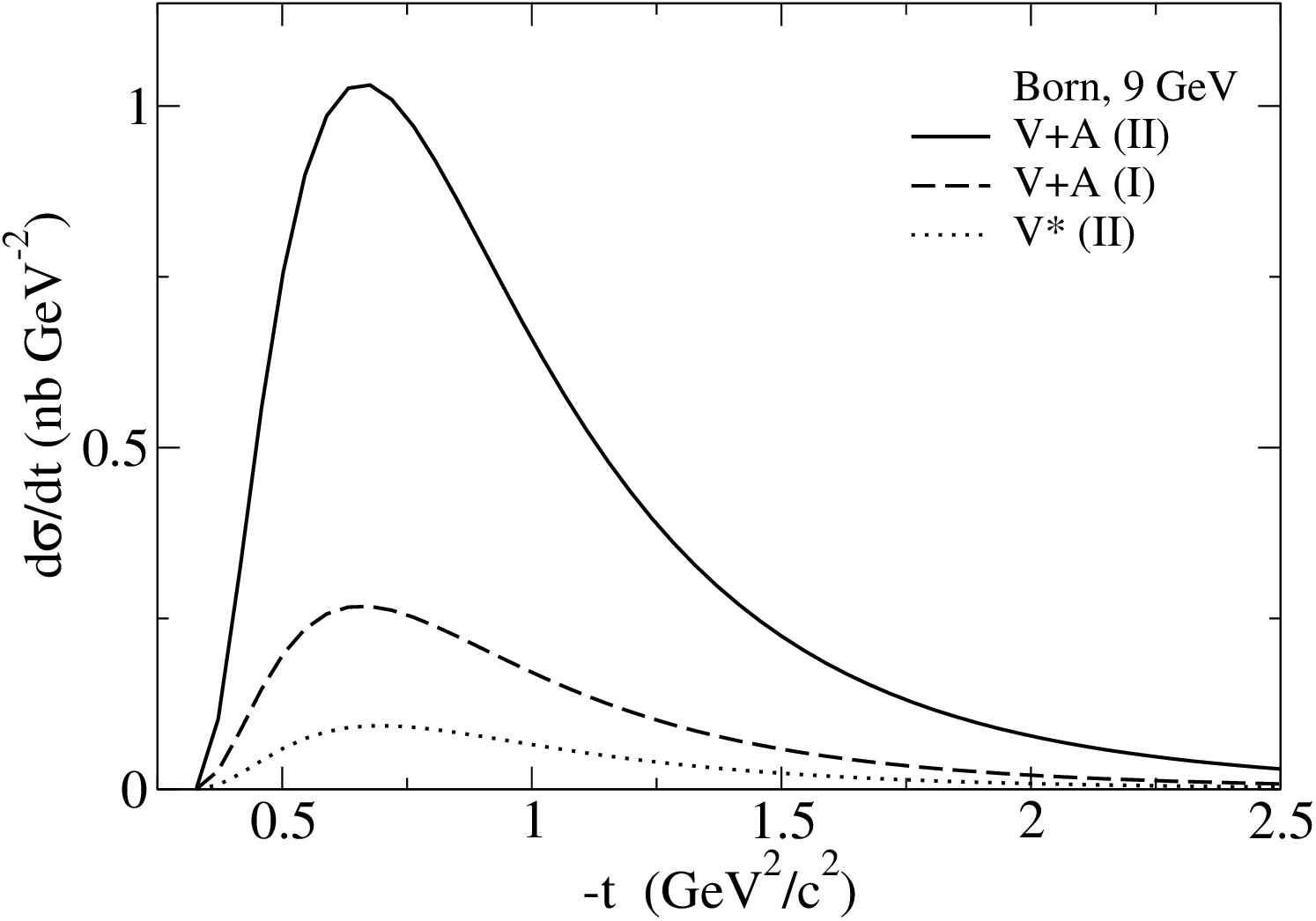,scale=0.2}

$(a)$\hspace{55mm}$(b)$\hspace{55mm}$(c)$
\vspace*{-.25cm}
\caption{The $\Lambda\bar\Lambda$ photoproduction for $E_\gamma=\,$9 GeV. 
The same content of panels as in Fig.~\ref{f2}.}
\label{f4} 
\end{center}

\begin{center}
\vspace*{.15cm}
\epsfig{figure=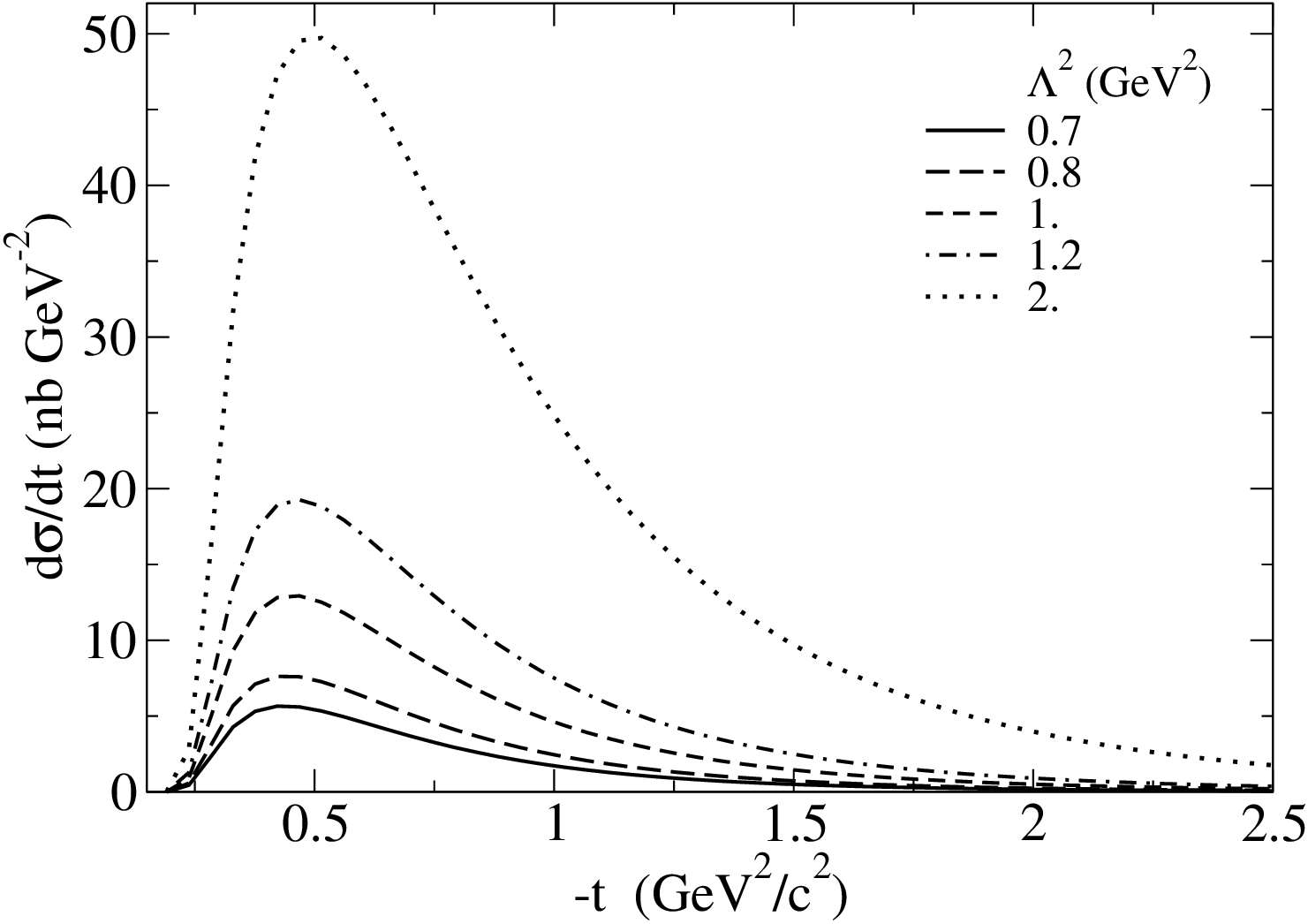,scale=0.2}
\vspace*{-.25cm}
\caption{The $p\bar p$ photoproduction for $E_\gamma=\,$9 GeV. 
The Born approximation for different values of the cutoff parameters 
$\Lambda = \Lambda_V = \Lambda_A$ ($\Lambda^2=\,$0.8, 1, 1.2, and 2 GeV$^2$) 
is shown in comparison to the choice $\Lambda^2=\,$0.7 GeV$^2$ (solid curve) 
used in this work.}
\label{f5}
\end{center}
\end{figure}

\end{document}